\documentclass[11pt]{article}

\usepackage{amsmath,amssymb,graphicx,color,amsthm}
\usepackage[linesnumbered,ruled]{algorithm2e}
\usepackage{tikz}
\usepackage{dsfont}
\usepackage{mathtools}
\usetikzlibrary{shapes.geometric, arrows}
\usepackage{algorithmic}
\usepackage[english]{babel}
\usepackage[utf8]{inputenc}
\usepackage{mathrsfs}  
\usepackage{afterpage}
\usepackage{mathtools}
\usepackage{amsfonts}
\usepackage[colorinlistoftodos]{todonotes}
\usepackage{enumerate}
\usepackage[round]{natbib}
\usepackage{hyperref}
\newcommand{\excise}[1]{}

\usepackage{newfloat}
\DeclareFloatingEnvironment[name={Supplementary Fig}]{suppfigure}

\renewcommand\iff{\Leftrightarrow}

\newtheorem*{example*}{Example}

\DeclareMathOperator\argmax{argmax}

\DeclarePairedDelimiterX{\infdivx}[2]{(}{)}{%
	#1\;\delimsize\|\;#2%
}

\newcommand{\RNum}[1]{\uppercase\expandafter{\romannumeral #1\relax}}

\begin{document}%%%%%%%%%%%%%%%%%%%%%%%%%%%%%%%%%%%%%%%%%%%%%%%%%%%%%%
	%%%%%%%%%%%%%%%%%%%%%%%%%%%%%%%%%%%%%%%%%%%%%%%%%%%%%%%%%%%%%%%%%%%%%%

\title{Contrastive latent variable modeling with application to case-control sequencing experiments}
\author{\\[1ex]Andrew Jones$^{1}$, F. William Townes$^1$, Didong Li$^{1,2}$, and Barbara E. Engelhardt$^{1,3}$\\
{\em $^{1}$Department of Computer Science, Princeton University}\\
{\em $^{2}$Department of Biostatistics, University of California, Los Angeles}\\
{\em $^{3}$Center for Statistics and Machine Learning, Princeton University}

}
\maketitle

\begin{abstract}
\noindent High-throughput RNA-sequencing (RNA-seq) technologies are powerful tools for understanding cellular state. Often it is of interest to quantify and summarize changes in cell state that occur between experimental or biological conditions. Differential expression is typically assessed using univariate tests to measure gene-wise shifts in expression. However, these methods largely ignore changes in transcriptional correlation. Furthermore, there is a need to identify the low-dimensional structure of the gene expression shift to identify collections of genes that change between conditions. Here, we propose contrastive latent variable models designed for count data to create a richer portrait of differential expression in sequencing data. These models disentangle the sources of transcriptional variation in different conditions, in the context of an explicit model of variation at baseline. Moreover, we develop a model-based hypothesis testing framework that can test for global and gene subset-specific changes in expression. We test our model through extensive simulations and analyses with count-based gene expression data from perturbation and observational sequencing experiments. We find that our methods can effectively summarize and quantify complex transcriptional changes in case-control experimental sequencing data.
\end{abstract}

\section{Introduction}
High-throughput RNA-sequencing technologies have emerged as useful tools for understanding transcriptional patterns. Recently, single-cell RNA-sequencing (scRNA-seq) technologies have allowed for investigation of these patterns at the level of individual cells. Together, these sequencing technologies have revealed new insights about a range of biological questions, from how cell types differ from one another to how cells respond to therapeutic drugs. In many scRNA-seq and bulk RNA-seq experiments, there are gene expression readouts for two or more experimental conditions or biological traits, such as tumor versus normal~\citep{young2018single,kinker2020pan}, drug exposure versus placebo exposure~\citep{srivastava2010testing,mcfarland2020multiplexed}, or ventilated versus non-ventilated lung tissue~\citep{gtex2020gtex}. In these cases, it is of interest to understand how the transcription levels differ in a \emph{foreground dataset} (collected from the treatment condition) relative to a \emph{background dataset} (collected from the control condition). These changes are traditionally identified using methods for differential gene expression, which estimate the average shift in expression levels between conditions.

Most differential expression methods for RNA-seq and scRNA-seq compute the univariate change between conditions for each gene separately~\citep{kharchenko2014bayesian,finak2015mast,qiu2017single}. In scRNA-seq, this amounts to treating each cell as an independent sample from one of two distributions over expression state; then each gene is considered marginally. These analyses use the repeated samples to quantify the uncertainty in the estimate of differential expression.

However, these methods for differential expression ignore a fundamental benefit of collecting scRNA-seq data over bulk RNA-sequencing: The ability to quantify population-level variation within a single sample across all of the cells. 
Population variation exists in pools of single cells, even from the same tissue. This gives us an opportunity to find differences in variation among genes, not just each gene itself.
Furthermore, it is believed that these transcriptional changes follow a low-dimensional structure.~\citep{dixit2016perturb,mcfarland2020multiplexed,becht2019dimensionality}. In particular, cells' fixed energy budgets and strongly correlated and interacting gene networks constrain the types of structural changes in gene expression between conditions.
In other words, changes in expression levels can be described in fewer dimensions than there are genes --- a phenomenon that most differential expression methods fail to capture. This low-dimensional projection can then be studied to find groups of genes that change similarly~\citep{xia2015testing}, and to quantify gene covariation in a low-dimensional representation~\citep{townes2019feature,ding2018interpretable}. There is a need for methods that can robustly identify the low-dimensional structure of variation in gene expression, and how this structure differs across experimental and biological conditions.

To address this gap in methodology, we develop a family of probabilistic models --- contrastive Poisson latent variable models (CPLVMs) --- that are designed to estimate the low-dimensional structure of the transcriptional response to perturbations as measured using sequencing technologies. Existing contrastive dimension reduction methods have shown promise for understanding the global shifts in variation between multiple conditions~\citep{zou2013contrastive,severson2019unsupervised,abid2018exploring}. But these methods fall short in a number of ways. First, they typically assume normally-distributed data, and do not treat count data from sequencing methods appropriately. Second, these methods are not designed to properly normalize count-based sequencing profiles. Finally, these approaches do not provide a hypothesis testing framework for testing differential expression across multiple genes.

Our contrastive Poisson latent variable model (CPLVM) bridges the gaps between existing contrastive methods and the needs of sequencing experiments by addressing these issues. We use a Poisson data likelihood that accounts for the count-based data produced by sequencing technologies. We show that these methods can identify changes in experimental and biological conditions that standard differential expression methods are not able to detect. Using our model, we decompose two-condition scRNA-seq and RNA-seq data into a small set of interpretable, nonnegative factors. Moreover, we build a hypothesis testing framework that accommodates hypotheses at varying scales, from testing for global shifts in gene expression across conditions to testing for correlated changes to a small set of candidate genes.

In this paper, we first describe the CPLVM in the context of related work and motivated by experiments in scRNA-seq. We then demonstrate the utility of our model through extensive simulations and experiments with multiple gene expression datasets. Using case-control scRNA-seq readouts of cells exposed to genetic and chemical perturbations~\citep{dixit2016perturb,mcfarland2020multiplexed}, we show that the CPLVM can identify structure that is specific to the foreground data. Furthermore, we show that these methods can identify changes across case-control data that standard differential expression methods are not able to detect. In addition, we apply the CPLVM to RNA-seq measurements from coronary artery tissue of donors with heart disease and healthy donors~\citep{gtex2020gtex}. We also show that the CPLVM hypothesis testing framework identifies experiments in which a specific, structured change in a small group of genes has been observed.

\section{Methods}
\subsection{Problem definition}
We motivate the problem using a generic scRNA-seq experiment, but our models can be applied to other count-based sequencing technologies as well. A scRNA-seq experiment with two conditions yields a set of unique molecular identifier (UMI) counts for cells from each condition. In this paper, we call measurements from the control condition the \emph{background data} and measurements from the treatment condition the \emph{foreground data}.

Suppose there are $n$ cells measured in the background condition, and $m$ cells measured in the foreground condition, with gene expression measured across $p$ (total) genes. We denote the data in matrix form as $\mathbf{Y} \in  \mathbb{N}_0^{p \times n}$ and $\mathbf{X} \in  \mathbb{N}_0^{p \times m}$, which contain the UMI counts for each cell and gene in the background and foreground data, respectively. 

The column vectors $\mathbf{y}_i \in  \mathbb{N}_0^p$ and $\mathbf{x}_j \in \mathbb{N}_0^p$ denote UMI counts across the $p$ genes for cell $i = \{1,\dots,n\}$ or $j = \{1,\dots,m\}$ from their respective conditions.

In this study, we are concerned with characterizing the transcriptional structure that exists in the foreground data $\mathbf{X}$ but not in $\mathbf{Y}$, as well as identifying the structure that is shared between the conditions. Decomposing these sources of variation into interpretable, low-dimensional structure is critical to understanding the effects of a treatment or different biological condition on cell state, regulation, and dynamics. 

\subsection{Related work}
Several families of methods have been developed to characterize the changes in gene expression between experimental or biological conditions. In this section, we outline several of these approaches.

\subsubsection{Differential expression methods}
The most common approaches for quantifying transcriptional changes in bulk and scRNA-seq data are differential expression methods. In general, these approaches compute univariate estimates of the change in expression for each gene between conditions. We review several approaches below; see~\citet{wang2019comparative} for a thorough review and benchmarking of differential expression methods in the context of scRNA-seq.

Most differential expression methods use linear models or generalized linear models (GLMs) to estimate the change in gene expression. For example, Multi-Input Multi-Output Single-Cell Analysis (MIMOSCA) --- which was developed specifically for the setting of genetic perturbation experiments in scRNA-seq --- uses a linear model with a Gaussian noise assumption~\citep{dixit2016perturb}. Specifically, it assumes the following model for the log-transformed and normalized UMI counts:
\begin{align}
    \log_2\left(\frac{\mathbf{y}_i}{n_i} s + c\right) &= \boldsymbol{\beta}_0 + \boldsymbol{\epsilon} \label{eq:linearmodel_log1} \\
    \log_2\left(\frac{\mathbf{x}_j}{n_j} s + c\right) &= \boldsymbol{\beta}_1 + \boldsymbol{\beta}_0 + \boldsymbol{\epsilon} \label{eq:linearmodel_log2} \\
    \boldsymbol{\epsilon} &\sim \mathcal{N}(\mathbf{0}, \sigma^2 I) \label{eq:linearmodel_log3}
\end{align}
where $n_i=\sum_k y_{ik}$ is the total number of counts in cell $i$ ($n_j$ is similarly defined), $s$ is a constant multiplicative factor, and $c$ is a ``pseudocount'' added to avoid taking $\log_2(0)$ (typically, $c=1$). The coefficient vector $\boldsymbol{\beta}_1 \in \mathbb{R}^p$ then captures the average fold-change in gene expression for a single gene between the conditions, and can be tested directly for significance.

GLMs more flexibly capture non-Gaussian data likelihoods. For example, in the context of scRNA-seq count data, a popular choice is the Poisson likelihood. In this case, the UMI count for gene $k$ in each cell is assumed to be a draw from a Poisson distribution, whose rate parameter is a transformation of the linear predictor:
\begin{align}
    y_{ik} &\sim \text{Poisson}\left(\frac{n_i}{s}  g^{-1}(\beta_{0k})\right) \;\;\; i=1, \dots, n \label{eq:poisson_glm1} \\
    x_{jk} &\sim \text{Poisson}\left(\frac{n_j}{s}  g^{-1}(\beta_{0k} + \beta_{1k})\right) \;\;\; j=1, \dots, m. \label{eq:poisson_glm2}
\end{align}
The canonical link function $g(\cdot)$ for a Poisson likelihood, $\log(\cdot)$, is typically used in the Poisson setting. Several existing methods, such as single-cell differential expression (SCDE~\citealt{kharchenko2014bayesian}), use a Poisson GLM to identify differential expression across conditions in scRNA-seq data. Closely related to the Poisson GLM, a common approach is to allow for overdispersion by using a negative binomial likelihood, which is equivalent to a gamma-Poisson mixture~\citep{love2014moderated,robinson2010edger,hafemeister2019normalization}. One method uses a zero-inflated negative binomial to model dropout events~\citep{miao2018desingle}.

Other distributional assumptions have also been proposed for differential expression in sequencing data. One approach, Model-based Analysis of Single-cell Transcriptomics (MAST), uses a hurdle model combined with a Gaussian likelihood~\citep{finak2015mast}. Another approach, scDD, uses a Dirichlet process mixture of Gaussians to model the potentially multimodal expression across cells and computes Bayes factors to quantify differential expression~\citep{korthauer2016statistical}. Other nonparametric approaches have also been proposed, including using Earth Mover's Distance to quantify expression changes~\citep{nabavi2016emdomics} and Cramér-von Mises and Kolmogorov-Smirnov hypothesis tests~\citep{delmans2016discrete}.

While differential expression methods have proven to be reliable for identifying changes that occur in individual genes across conditions, they typically ignore any correlation structure between genes. This is an important limitation, as gene expression has been shown to have substantial correlation structure~\citep{stuart2003gene}, and identifying changes in this structure across conditions is of great interest.

\subsubsection{Two-sample covariance comparison methods}
Another related line of work has focused on identifying differences in the covariance between two conditions. Most commonly, these approaches rely on a hypothesis test to decide whether covariance matrices $\Sigma_X, \Sigma_Y$ are different:
\begin{equation*}
    H_0: ~~\Sigma_X = \Sigma_Y, ~~~~~ H_1: ~~\Sigma_X \neq \Sigma_Y.
\end{equation*}

There exist a number of such tests in the setting of low-dimensional data, including modified multivariate generalizations of Levene's test~\citep{o1992robust} and the commonly-used likelihood ratio test~\citep{lrt}. But high-dimensional data pose a greater challenge. Since these existing tests were designed for small numbers of features $p$ relative to sample size $n$, they have poor statistical power when applied to high-dimensional data where $p \gg n$, and in some cases are not even well-defined \citep{cai2013two}.

Some covariance inequality tests have been designed to address the problem of high-dimensional data by using estimators of the distance between covariance matrices based on the Frobenius norm \citep{li2012two,srivastava2010testing}. However, these tests have low power to detect the true effect when the differences between the covariance matrices are sparse.

Two techniques have emerged to address both the issue of high-dimensional data and the possibility of a small number of entries driving the differences between two covariance matrices. The first approach uses a test statistic based on the largest standardized difference between the two covariance matrices' entries~\citep{cai2013two}. The second approach uses a Gaussian graphical model framework to infer the differential network structure~\citep{xia2015testing}. Other related approaches consider building Gaussian graphical models from precision matrices and identifying network edges that are differentially identified across two conditions~\citep{glass2013passing}. 

More recently, a hypothesis testing approach that is robust in the setting of high-dimensionality and low sample size was developed using the strongly spiked eigenvalue (SSE) model~\citep{aoshima2018two,ishii2019equality}. The SSE model assumes the first eigenvalues $\lambda_{1x}, \lambda_{1y}$ of the covariance matrices $\Sigma_x, \Sigma_y$ are ``strongly spiked'' relative to the subsequent eigenvalues, in the sense of
\begin{equation*}
    \liminf_{p \to \infty} \left\{ \frac{\lambda_{1i}}{\text{tr}(\Sigma_i^2)} \right\} > 0.
\end{equation*}
The authors show that the SSE assumption is reasonable in many high-dimensional settings, especially when $p\gg n$. They derived the limiting distributions for test statistics under on this model, as well as the power and size of the accompanying hypothesis tests. They found that the SSE model had greater statistical power compared to previous models that assumed more diffuse eigenvalue spectra. However, methods for hypothesis testing neglect the goals of exploratory data analysis, including identifying interpretable, low-dimensional factors that explain the changes in covariance structure across conditions.

\subsubsection{Contrastive dimension reduction methods}
Contrastive dimension reduction methods estimate low-dimensional changes in variation between conditions. In particular, these methods aim to identify variation that exists in the foreground data but not in the background data. Furthermore, they typically assume that the variation in each condition can be explained by a small number of latent dimensions.

As one of the first steps in this direction, a framework for contrastive learning in mixture models was proposed~\citep{zou2013contrastive}. This approach assumes that the background and foreground data are generated from a set of mixture distributions, some of which are shared between the two conditions, and some of which are exclusive to each condition. Specifically, given a set of mixture parameters $\{\mu_\ell\}, w_\ell\}_{\ell=1}^L$, condition-specific mixture weights $\{w_\ell^\text{b}\}_{\ell=1}^L, \{w_\ell^\text{f}\}_{\ell=1}^L$, and three disjoint index sets $A, B, C \subseteq [L]$, it assumes $\mathbf{Y}$ and $\mathbf{X}$ are drawn from a set of mixtures:
\begin{align*}
    p(\mathbf{y}_i; \{\mu_\ell, w_\ell\}_{\ell \in A \cup B}) &= \sum\limits_{\ell \in A \cup B} w_\ell^\text{b} f(\mathbf{y}_i | \mu_\ell), \;\;\;\; i = 1, \dots, n \\
    p(\mathbf{x}_j; \{\mu_\ell, w_\ell\}_{\ell \in B \cup C}) &= \sum\limits_{\ell \in B \cup C} w_\ell^\text{f} f(\mathbf{x}_j | \mu_\ell), \;\;\;\; j = 1, \dots, m.
\end{align*}
Note that the mixture components indexed by $B$ are shared between the conditions, while those indexed by $A$ and $C$ are unique to the background and foreground, respectively. This framework encompasses several general models, including topic models, such as Latent Dirichlet Allocation (LDA). The authors were primarily interested in estimating the foreground-specific model parameters, $\{\mu_\ell, w_\ell^\text{f}\}_{\ell \in A}$. Their inference approach relies on a tensor decomposition method for estimating the foreground-specific latent components without estimating the background or shared components.

As an important special case of this contrastive learning framework, contrastive principal component analysis (CPCA) was derived explicitly, which extends the classical PCA method~\citep{abid2018exploring}. Specifically, given the sample covariance matrices of the background and foreground conditions, $\widehat{\Sigma}_x$ and $\widehat{\Sigma}_y$, the objective function of CPCA seeks to find a unit vector $\mathbf{v}$ that maximizes the variance in the foreground and minimizes the variance in the background:
\begin{equation*}
    \argmax_{\mathbf{v} \in \mathbb{S}^{p-1}} \left\{\mathbf{v}^\top \widehat{\Sigma}_x \mathbf{v} - \gamma \mathbf{v}^\top \widehat{\Sigma}_y \mathbf{v}\right\}.
\end{equation*}
Here, $\gamma \geq 0$ is a tuning parameter controlling the relative influence of the background data. When $\gamma=0$, this model reduces to PCA on the foreground data. This problem can be solved analytically: the top $k$ ``contrastive principal components'' correspond to the $k$ eigenvectors $\begin{bmatrix} \mathbf{u}_1, \dots, \mathbf{u}_k \end{bmatrix}$, which represent the top $k$ eigenvalues $\lambda_1 \geq \cdots \geq \lambda_k$ of the differential covariance:
\begin{equation*}
C = \widehat{\Sigma}_x - \gamma \widehat{\Sigma}_y.
\end{equation*}
The authors showed that CPCA accurately recovers structure that is unique to the foreground data, and CPCA is able to identify heterogeneous responses in two-condition gene expression data~\citep{abid2017contrastive,abid2018exploring}.

A sparse version of CPCA was recently developed, which allows for greater interpretability of the estimated components, especially in high-dimensional settings~\citep{boileau2020exploring}. Building off of sparse PCA~\citep{zou2006sparse}, which uses element-wise $\ell_1$ regularization to encourage zeros in the loadings matrix, the authors propose an estimation procedure that alternates between estimating the principal components and the sparse loadings matrix. They demonstrate the behavior of sparse CPCA on a series of gene and protein expression datasets.

Most closely related to our work, probabilistic counterparts to CPCA have been proposed. The contrastive latent variable model (CLVM, \citealt{severson2019unsupervised}) captures structure that is unique to the foreground data, as well as structure shared between the conditions. In particular, the shared variation is described by a set of latent variables $\{\mathbf{z}_i^{\text{b}}\}_{i=1}^n$ and $\{\mathbf{z}_j^{\text{f}}\}_{j=1}^m$, and the foreground-specific variation is captured by another set of latent variables $\{\mathbf{t}_j\}_{j=1}^m$. Using Gaussian likelihoods and priors, the CLVM has the following form:
\begin{align*}
    \mathbf{y}_i | \mathbf{z}_i^{\text{b}} &\sim \mathcal{N}(\mathbf{S} \mathbf{z}_i^{\text{b}} + \boldsymbol{\mu}^{\text{b}}, \sigma^2 I) \\
    \mathbf{x}_j | \mathbf{z}_j^{\text{f}}, \mathbf{t}_j &\sim \mathcal{N}(\mathbf{S} \mathbf{z}_j^{\text{f}} + \mathbf{W} \mathbf{t}_j + \boldsymbol{\mu}^{\text{f}}, \sigma^2 I) \\
    \mathbf{z}_i^{\text{b}} \sim \mathcal{N}(\mathbf{0}, I), \;\;\; \mathbf{z}_j^{\text{f}} &\sim \mathcal{N}(\mathbf{0}, I), \;\;\; \mathbf{t}_j \sim \mathcal{N}(\mathbf{0}, I).
\end{align*}
Here, $\mathbf{S} \in \mathbb{R}^{p \times k_1}$ and $\mathbf{W} \in \mathbb{R}^{p \times k_2}$ are loadings matrices that map from the latent dimensions $k_1$ and $k_2$ to the data feature dimension $p$. Through experiments with gene expression and image data, the authors showed that the CLVM can disentangle low-dimensional latent structure that is shared between the two conditions and structure that is specific to the foreground.

Another model-based contrastive method, probabilistic contrastive principal component analysis (PCPCA, \citealt{li2020probabilistic}) was developed as a direct generalization of CPCA and probabilistic PCA. PCPCA provides a simple estimation procedure based on a likelihood ratio and was shown to be robust to noise and missing data. In applications, PCPCA was successful in identifying subgroup structure in case-control gene expression data. Although these probabilistic models have many advantages over previous approaches, both CLVM and PCPCA assume Gaussian errors, which is not ideal for modeling count-based expression profiles.

While contrastive dimension reduction methods have shown promise for analyzing two-condition datasets, there remains a need to adapt these methods to the setting of sequencing data, where observations are counts of RNA sequence fragments mapping to genes across the genome. Moreover, there is a substantial need to provide a common framework for both factor analysis and hypothesis testing in these models when it is useful to quantify the statistical significance of changes in the covariance structure of expression across cases and controls in an experimental setting.

\section{Contrastive Poisson latent variable models for scRNA-seq}
In this study, we develop a family of contrastive Poisson latent variable models (CPLVMs) that are designed to capture variation and covariation among count data that are unique to the foreground condition, as well as variation and covariation that are shared between the foreground and background. Furthermore, we build a hypothesis testing framework that quantifies support for structured changes in variation across conditions. Throughout, we rely on principled probabilistic modeling of count data, rather than data transformations and Gaussian models. In the context of sequencing data, our model explicitly accounts for the count-based nature of expression profiles, while decomposing case-control data into a small set of interpretable factors.

In the following sections, we first describe the CPLVM. Then we explain our inference procedure for the CPLVM, and we develop the corresponding hypothesis testing framework.

\subsection{CPLVM definition}
As above, let $\mathbf{Y} \in \mathbb{N}_0^{p \times n}$ and $\mathbf{X} \in \mathbb{N}_0^{p \times m}$ be the count matrices for $p$ genes and $n$, $m$ cells for the background and foreground conditions, respectively.

The CPLVM assumes that transcription variation in a sequencing experiment with multiple conditions can be described by a small set of latent factors. In particular, the model assumes that the variation shared between the conditions is described by a set of $k_1$-dimensional latent variables, $\{\mathbf{z}_i^{\text{b}}\}_{i=1}^n$ and $\{\mathbf{z}_j^{\text{f}}\}_{j=1}^m$. Furthermore, we assume that the foreground-specific variation is captured by another set of $k_2$-dimensional latent variables $\{\mathbf{t}_j\}_{j=1}^m$. To describe the mapping between these latent spaces and the data, we introduce loadings matrices $\mathbf{S} \in \mathbb{R}_+^{k_1 \times p}$ and $\mathbf{W} \in \mathbb{R}_+^{k_2 \times p}$, which map to the data space of dimension $p$ from the shared latent space of dimension $k_1$ and the foreground-specific latent space of dimension $k_2$.

To account for varying numbers of total counts between cells and experimental conditions, we include size factors $\alpha_i^{\text{b}}$ and $\alpha_j^{\text{f}}$ for each cell in each condition; these terms control for technical variation in the total number of counts per cell. Furthermore, to account for shifts in each gene's mean counts between conditions, we also include gene-specific multiplicative scale parameters $\boldsymbol{\delta} \in \mathbb{R}_+^p$. 
These terms are analogous to the gene-wise additive intercept terms in a linear model, but we constrain them to be nonnegative in this model. Thus, for gene $k$, $0<\widehat{\delta}_k<1$ indicates that there is lower relative expression of gene $k$ in the background cells, while $\widehat{\delta}_k>1$ indicates higher relative expression of gene $k$ in the background cells.

The full generative model for the nonnegative CPLVM is then
\begin{align}
    \mathbf{y}_i | \mathbf{z}_i & \sim \text{Poisson}\left(\alpha_i^{\text{b}} \boldsymbol{\delta} \odot \left(\mathbf{S}^\top \mathbf{z}_i^{\text{b}}\right)\right) \label{eq:cplvm1} \\
    \mathbf{x}_j | \mathbf{z}_j, \mathbf{t}_j & \sim \text{Poisson}\left(\alpha_j^{\text{f}} \left( \mathbf{S}^\top \mathbf{z}_j^{\text{f}} + \mathbf{W}^\top \mathbf{t}_j \right) \right) \label{eq:cplvm2} \\
    z_{il}^{\text{b}} \sim \text{Gamma}(\gamma_1, \beta_1)&, \;\;\; z_{jl}^{\text{f}} \sim \text{Gamma}(\gamma_2, \beta_2), \;\;\; t_{jd} \sim \text{Gamma}(\gamma_3, \beta_3), \label{eq:cplvm3} \\
    W_{kd} \sim \text{Gamma}(\gamma_4, \beta_4)&, \;\;\; S_{jl} \sim \text{Gamma}(\gamma_5, \beta_5) \label{eq:cplvm4}, \;\;\; \boldsymbol{\delta} \sim \text{LogNormal}(0, \textbf{I}_p),
\end{align}
where $l \in \{1, \dots, k_1\}$, $d \in \{1, \dots, k_2\}$, and $\odot$ represents a Hadamard (element-wise) product. Following previous work~\citep{lopez2018deep}, we place log normal priors on the size factors $\alpha_i, \alpha_j$ with parameters given by the empirical mean and variance of the log total counts for each cell.

\subsection{Stochastic variational inference for the CPLVM}
For a given experiment, we are interested in estimating the posterior distribution of $\mathbf{Z}^{\text{b}}, \mathbf{Z}^{\text{f}}, \mathbf{T}, \mathbf{S}$, and  $\mathbf{W}$ given the data $\{\mathbf{Y}, \mathbf{X}\}$. Since the true posterior is intractable, we use a variational approximation.

Specifically, we perform approximate posterior inference on the latent variables $\mathbf{Z}^{\text{b}}, \mathbf{Z}^{\text{f}}, \mathbf{T}$, the loadings matrices $\mathbf{S}, \mathbf{W}$, and the mean-shift parameter $\boldsymbol{\delta}$ using a mean-field variational approximation. In other words, we approximate the true posterior distribution $p(\mathbf{Z}^{\text{b}}, \mathbf{Z}^{\text{f}}, \mathbf{T}, \mathbf{S}, \mathbf{W}, \boldsymbol{\delta} | \mathbf{Y}, \mathbf{X})$ with a variational posterior distribution $q$ that fully factorizes: 
\begin{equation*}
    q(\mathbf{Z}^{\text{b}}, \mathbf{Z}^{\text{f}}, \mathbf{T}, \mathbf{S}, \mathbf{W}) = q(\mathbf{Z}^{\text{b}}) q(\mathbf{Z}^{\text{f}}) q(\mathbf{T}) q(\mathbf{S}) q(\mathbf{W}) q(\boldsymbol{\delta}).
\end{equation*}

For numerical stability and speed, we specify each of these variational distributions to be log normal for the CPLVM:
\begin{align*}
    q(\mathbf{z}_i) = \text{LogNormal}(\boldsymbol{\mu}_1, \sigma_1^2 I_{k_1}), \;\; q(\mathbf{z}_j) &= \text{LogNormal}(\boldsymbol{\mu}_2, \sigma_2^2 I_{k_1}), \;\; q(\mathbf{t}_j) = \text{LogNormal}(\boldsymbol{\mu}_3, \sigma_3^2 I_{k_2}), \\
    q(\mathbf{S}_l) = \text{LogNormal}(\boldsymbol{\mu}_4, \sigma_4^2 I_p), \;\; q(\mathbf{W}_d) &= \text{LogNormal}(\boldsymbol{\mu}_5, \sigma_5^2 I_p), \;\; q(\boldsymbol{\delta}) = \text{LogNormal}(\boldsymbol{\mu}_6, \sigma_6^2 I_p).
\end{align*}

We perform approximate inference by minimizing the Kullback-Leibler (KL) divergence between the true posterior and the approximate posterior with respect to the variational parameters. This is equivalent to maximizing the lower bound on the log marginal likelihood of the data, known as the evidence lower bound (ELBO):
\begin{equation}\label{eqn:ELBO}
    \log p(\mathbf{Y}, \mathbf{X}) \geq \mathbb{E}_{\mathcal{Z} \sim q(\mathcal{Z})} \left[ \log \frac{q(\mathcal{Z})}{p(\mathbf{Y}, \mathbf{X}, \mathcal{Z})} \right],
\end{equation}
where $\mathcal{Z} = \{\mathbf{Z}^{\text{b}}, \mathbf{Z}^{\text{f}}, \mathbf{T}, \mathbf{S}, \mathbf{W}, \boldsymbol{\delta}\}$.

We use stochastic gradient descent to minimize the negative ELBO~\citep{hoffman2013stochastic}. We define and fit the variational model using TensorFlow probability~\citep{dillon2017tensorflow}.

\subsection{Contrastive generalized latent variable model}
We also develop a second CPLVM that allows the factors to be negative by leveraging the exponential family distributions. In this case, we use a $\log$-link function to transform the linear predictors to $\mathbb{R}_+$, similar to a generalized linear model (GLM). We call this model the contrastive generalized latent variable model (CGLVM). Here, in place of the multiplicative scale terms $\boldsymbol{\delta}$ in the CPLVM, we use additive coefficient terms $\boldsymbol{\mu}^{\text{f}}, \boldsymbol{\mu}^{\text{b}} \in \mathbb{R}^p$, similar to a traditional GLM.
\begin{align}
    \mathbf{y}_i | \mathbf{z}_i & \sim \text{Poisson}\left(\exp \left\{\left(\mathbf{S}^\top \mathbf{z}_i^{\text{b}} + \boldsymbol{\mu}^{\text{f}} + \log \alpha_i^{\text{b}} \right) \right \} \right) \label{eq:cplvm_glm1} \\
    \mathbf{x}_j | \mathbf{z}_j, \mathbf{t}_j & \sim \text{Poisson}\left(\exp \left\{\left( \mathbf{S}^\top \mathbf{z}_j^{\text{f}} + \mathbf{W}^\top \mathbf{t}_j + \boldsymbol{\mu}^{\text{b}} + \log \alpha_j^{\text{f}} \right) \right\} \right) \label{eq:cplvm_glm2} \\
    \mathbf{z}_i^{\text{b}} &\sim \mathcal{N}(\mathbf{0}, I), \;\;\; \mathbf{z}_j^{\text{f}} \sim \mathcal{N}(\mathbf{0}, I), \;\;\; \mathbf{t}_j \sim \mathcal{N}(\mathbf{0}, I) \label{eq:cplvm_glm3} \\
    \mathbf{W}_d &\sim \mathcal{N}(0, \textbf{I}_p), \;\;\; \mathbf{S}_l \sim \mathcal{N}(0, \textbf{I}_p), \;\;\; \boldsymbol{\mu}^{\text{b}}, \boldsymbol{\mu}^{\text{f}} \sim \mathcal{N}(0, \textbf{I}_p) \label{eq:cplvm_glm4}
\end{align}
where $l \in \{1, \dots, k_1\}$, $d \in \{1, \dots, k_2\}$. We place log normal priors on the size factors $\alpha_i^\text{b}, \alpha_j^\text{f}$, similar to those for the CPLVM.

For inference in the CGLVM, we again use a variational approach. Here, we specify the variational distributions as multivariate Gaussians:
\begin{align*}
    q(\mathbf{z}_i) = \mathcal{N}(\boldsymbol{\mu}_1, \sigma_1^2 I_{k_1}), \;\; q(\mathbf{z}_j) &= \mathcal{N}(\boldsymbol{\mu}_2, \sigma_2^2 I_{k_1}), \;\; q(\mathbf{t}_j) = \mathcal{N}(\boldsymbol{\mu}_3, \sigma_3^2 I_{k_2}), \\
    q(\mathbf{S}_l) = \mathcal{N}(\boldsymbol{\mu}_4, \sigma_4^2 I_p), \;\; q(\mathbf{W}_d) &= \mathcal{N}(\boldsymbol{\mu}_5, \sigma_5^2 I_p), \;\; q(\boldsymbol{\mu}^{\text{b}}), q(\boldsymbol{\mu}^{\text{f}}) = \mathcal{N}(\boldsymbol{\mu}_6, \sigma_6^2 I_p).
\end{align*}
Similar to the CPLVM, we apply stochastic variational inference to optimize the ELBO.

% \vspace{4mm}
\begin{center}
\begin{table}[t!]
\begin{tabular}{ |c||c|c|c|c|c|c|  }
 \hline
  & Contrastive & LVM & Background & Orthogonal & Counts & Nonnegative  \\
 \hline
 \hline
 PCA & & & & x & &  \\
 \hline
 PPCA & & x & & x & &  \\
 \hline
 NMF & & & & & & x  \\
 \hline
 CPCA & x & & & x & &  \\
 \hline
 PCPCA & x & x & & x & &  \\
 \hline
 CLVM & x & x & x & & &  \\
 \hline
 CGLVM (ours) & x & x & x & & x &  \\
 \hline
 CPLVM (ours) & x & x & x & & x & x  \\
 \hline
\end{tabular}
\vspace{1mm}
\caption{\textbf{Overview of related dimension reduction methods.} \emph{Contrastive}: Method directly models the contrast between two datasets. \emph{LVM}: Method has a latent variable model formulation. \emph{Background}: Method includes an explicit model of of the background data. \emph{Orthogonal}: Method constrains factors to be orthogonal to one another. \emph{Counts}: Method directly accounts for count-based data. \emph{Nonnegative}: Method has nonnegative factors and loadings.}
\label{tab:table1}
\end{table}
\end{center}
% \vspace{3mm}

\subsection{Hypothesis testing with CPLVMs}
In addition to exploratory data analysis using the model defined above, the CPLVM framework allows us to test hypotheses about whether the transcriptional structure is altered between conditions. Specifically, these tests quantify the extent to which the model's goodness of fit is improved when including the foreground-specific latent variables in addition to the shared latent variables.

The Bayesian framework of our model allows for model comparison using Bayes factors~\citep{goodman1999toward}. Bayes factors compare the ratio of data log likelihoods between an alternative model $\mathcal{M}_1$ and a null model $\mathcal{M}_0$, integrating over model parameters. Specifically, the Bayes factor is the ratio of model evidence (or marginal likelihoods): 
\begin{equation*}
    \text{BF} = \frac{p(\mathbf{Y}, \mathbf{X} | \mathcal{M}_1)}{p(\mathbf{Y}, \mathbf{X} | \mathcal{M}_0)} = \frac{\int_{\Theta_1} p(\mathbf{Y}, \mathbf{X} | \theta_1, \mathcal{M}_1) p(\theta_1 | \mathcal{M}_1) d\theta_1}{\int_{\Theta_0} p(\mathbf{Y}, \mathbf{X} | \theta_0, \mathcal{M}_0) p(\theta_0 | \mathcal{M}_0) d\theta_0}.
\end{equation*}

In practice, the $\log$-Bayes factor is often used, and the null hypothesis is rejected if it surpasses some threshold $\tau$: 
\begin{equation*}
    \text{Reject $H_0$} \iff \log p(\mathbf{Y}, \mathbf{X} | \mathcal{M}_1) - \log p(\mathbf{Y}, \mathbf{X} | \mathcal{M}_0) > \tau.
\end{equation*}

Here, we implicitly assume equal prior weight on the null and alternative hypotheses, $p(\mathcal{M}_0)=p(\mathcal{M}_1)$, which we find to be well-calibrated in our numerical experiments.

Selecting a proper value for $\tau$ depends on the application area~\citep{kass1995bayes}. In practical settings, often $\tau$ is chosen based on a frequentist calibration of the hypothesis test.

In general, computing the model evidence requires solving an intractable integral, in turn making Bayes factors difficult to estimate. Following previous work~\citep{lopez2018deep}, we address this issue by approximating the model evidence with the ELBO (Equation \eqref{eqn:ELBO}), which is a lower bound on the evidence, leading to ELBO-based Bayes factors (EBFs). We note that the tightness of the lower bound on the evidence depends on a number of modeling choices --- for example, the choice of variational families and parameter initialization for stochastic VI --- and the gap between the ELBO and the evidence could be different for the numerator and the denominator. However, we find these EBFs to be reliable in a number of simulations for our models.
Thus, the general form of the CPLVM hypothesis test is the following:
\begin{equation}
    \text{Reject $H_0$} \iff \text{ELBO}_{\mathcal{M}_1} - \text{ELBO}_{\mathcal{M}_0} > \tau \label{eq:cplvm_bf}.
\end{equation}
Defining the null and alternative models depends on the hypothesis of interest. Here, we consider two types of hypotheses for our model: \emph{global} hypotheses and \emph{gene set} hypotheses.

\subsubsection{Global hypothesis test for changes in gene covariance structure}
In this paper, a global hypothesis test is one that considers changes in expression across all genes that occur between conditions. This type of test is useful for assessing the effect of interventions that are expected to impact the expression of many genes, and the covariance structure among those genes.

For these hypotheses, we propose constructing the null model by removing the latent variables specific to the foreground data $\mathbf{t}_j$. Hence, the null model for the global hypothesis is
\begin{align}
    \mathbf{y}_i | \mathbf{z}_i & \sim \text{Poisson}\left(\alpha_i^{\text{b}} \boldsymbol{\delta} \odot \mathbf{S} \mathbf{z}_i^{\text{b}}\right) \label{eq:H0_cplvm1} \\
    \mathbf{x}_j | \mathbf{z}_j & \sim \text{Poisson}\left(\alpha_j^{\text{f}} \left( \mathbf{S} \mathbf{z}_j^{\text{f}} \right) \right) \label{eq:H0_cplvm2} \\
    \mathbf{z}_i^{\text{b}} &\sim \text{Gamma}(\gamma, \beta_1), \;\;\; \mathbf{z}_j^{\text{f}} \sim \text{Gamma}(\gamma, \beta_2). \label{eq:H0_cplvm3}
\end{align}
Intuitively, the null model assumes that the latent structure in both matrices can be captured using a single, shared latent space $S$, and the samples from both matrices are projected onto that latent space using $\mathbf{z}_j^{\text{b}}$ and $\mathbf{z}_j^{\text{f}}$ for background and foreground data, respectively.

The alternative hypothesis is that there is structure specific to the foreground data across all genes. For the alternative model, we use the full CPLVM defined in Equations \eqref{eq:cplvm1}-\eqref{eq:cplvm4}. This model includes the latent variables specific to the foreground data $\mathbf{t}_j$.

\subsubsection{Gene-set hypothesis test for foreground changes to a subset of genes}
To test for changes in variation in foreground data relative to background data involving only a subset of genes, we propose a gene set hypothesis test. This test is useful to quantify support for changes to joint expression within specific known gene modules that are unique to the foreground matrix.

Suppose we would like to test for differential variation in a set of $G$ genes indexed by $\{\ell_1, \dots, \ell_G\}$. In this case, the null hypothesis is encoded in a model constructed by constraining rows of $\mathbf{W}$ indexed by $\{\ell_1, \dots, \ell_G\}$ to be zero. Intuitively, this model assumes no change in variation specific to that gene set in the foreground matrix relative to the background matrix.
To be precise, let $\mathbf{Q}$ be a $G \times p$ binary matrix which, when multiplied with $\mathbf{W}$, only takes rows corresponding to genes in the gene set. Specifically, for $g \in \{1, \dots, G\}$ and $k \in \{1, \dots, p\}$,
\begin{equation*}
    \mathbf{Q}_{gk} = \begin{cases} 1 & k=\ell_g \\
    0 & \text{otherwise}.
\end{cases}
\end{equation*}
Then the null model is
\begin{align}
    \mathbf{y}_i | \mathbf{z}_i & \sim \text{Poisson}\left(\alpha_i^{\text{b}} \boldsymbol{\delta} \odot \mathbf{S} \mathbf{z}_i^{\text{b}}\right) \label{eq:cplvm_geneset_H01} \\
    \mathbf{x}_j | \mathbf{z}_j & \sim \text{Poisson}\left(\alpha_j^{\text{f}} \left( \mathbf{S} \mathbf{z}_j^{\text{f}} + \mathbf{W} \mathbf{t}_j \right) \right) ~~ \text{s.t.}~~ \mathbf{Q} \mathbf{W} = \mathbf{0}_G, \label{eq:cplvm_geneset_H02}
\end{align}
where $\mathbf{0}_G$ is the zero vector of length $G$.
We specify the alternative hypothesis in a model that includes the full CPLVM as described in \eqref{eq:cplvm1}-\eqref{eq:cplvm4}.

A gene set hypothesis will test whether there are changes in the covariance structure specific to a subset of genes that are prespecified. To define gene sets of interest, one could use established gene set collections, such as the MSigDB sets~\citep{liberzon2015molecular}, as we show in our experiments below.

\section{Simulation results}
In this section, we evaluate the performance of the CGLVM and CPLVM using synthetic data. We compare the performance of these models to four related state-of-the-art methods for dimension reduction (PCA, NMF, CPCA, and PCPCA), a related linear model for differential expression discovery, and a two sample test for differential expression testing.

\subsection{Visualizing CGLVM and CPLVM latent spaces}
In order to demonstrate the behavior of the CGLVM and CPLVM, we first fit the model on simple two-dimensional simulated count data. In this dataset, the foreground data is made up of two subgroups, while the background data is homogeneous. Specifically, to generate count data with a prespecified covariance matrix $\Sigma$, we use a Gaussian copula with a Poisson likelihood. In particular, we generate the foreground samples $\{\mathbf{x}_i\}_{i=1}^n$ and background samples $\{\mathbf{y}_j\}_{j=1}^m$ as
\begin{align*}
    \mathbf{x}_i &= F^{-1}_\lambda(\Phi(z_i)) \\
    \mathbf{y}_j &= F^{-1}_\lambda(\Phi(z_j))
\end{align*}
where $F^{-1}_\lambda$ is the inverse CDF of a Poisson with parameter $\lambda$, $\Phi$ is the standard normal CDF, and $z_i, z_j \sim \mathcal{N}(0, \Sigma)$. We then shift the subgroups to give the background a mean of $\left(\begin{smallmatrix} 14 \\ 14 \end{smallmatrix}\right)$ and the foreground subgroups means of $\left(\begin{smallmatrix} 18 \\ 10 \end{smallmatrix}\right)$ and $\left(\begin{smallmatrix} 10 \\ 18 \end{smallmatrix}\right)$. Here, we set $\Sigma = \left(\begin{smallmatrix} 2.7 & 2.6 \\ 2.6 & 2.7 \end{smallmatrix}\right)$, $\lambda=10$, and $n=m=200$.

We fit the CGLVM and CPLVM, as well as the related methods listed above, on this dataset. For the CGLVM, we use a single latent dimension for the shared and foreground-specific compartments, $k_1=k_2=1$. For the CPLVM, we set $k_1=1, k_2=2$ in order to allow the nonnegative factors to discover negative associations.
% We then plot the one-dimensional subspace defined by each latent dimension of $\textbf{S}$ and $\textbf{W}$ (\autoref{figure1}e, f). For related methods, we plot their estimated factors in the same way (\autoref{figure1}a-d). 

\begin{figure}[t!]
\includegraphics[width=1.0\textwidth]{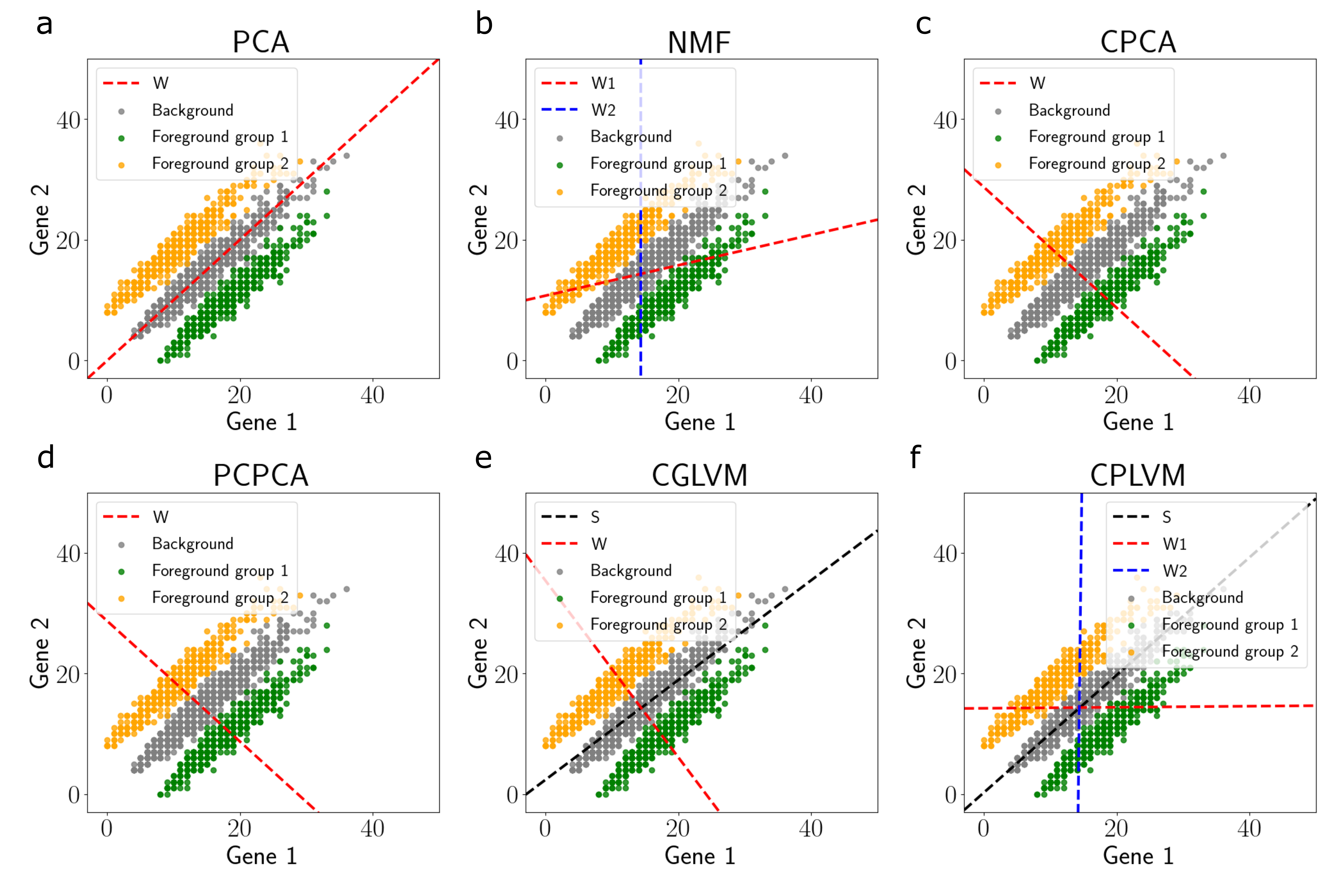}
\caption{\textbf{Illustration of the CPLVM with toy data.} Related dimension reduction methods applied to a toy dataset in which the foreground data contains two subgroups. (a) PCA ($k=1)$, (b) NMF ($k=2$), (c) CPCA ($k=1$), (d) PCPCA ($k=1$), (e) CGLVM (ours, $k_1=k_2=1$), (f) CPLVM (ours, $k_1=1, k_2=2$). In each, we plot the one-dimensional line defined by each column of the estimated loadings matrix from each method.
}
\label{figure1}
\end{figure}

We find that, in both the CGLVM and CPLVM, the subspaces defined by $\textbf{S}$ and $\textbf{W}$ picked up on the directions of shared and foreground-specific variation, respectively (\autoref{figure1}e, f). The other contrastive methods, CPCA and PCPCA, were able to detect the axis of variation unique to the foreground data, but these approaches do not have an explicit model for the background data (\autoref{figure1}c, d). Finally, PCA and NMF --- which do not model the contrast between the conditions --- are unable to identify the axis that separates the two foreground subgroups (\autoref{figure1}a, b).

This result suggests that the CPLVM is able to disentangle these two sources of variation: those that are shared between conditions, and those that are unique to the foreground.

\subsection{Discovering heterogeneous responses}
We next examined whether the CPLVM discovers the latent structure of a dataset in which there is a heterogeneous response across samples in the foreground data. To study this behavior, we generate a synthetic count dataset from the CPLVM model \eqref{eq:cplvm1}-\eqref{eq:cplvm4}. We set the parameters such that all background cells have the same latent state, but each foreground cell is drawn from one of two unique latent states. Specifically, we set the data dimension $p=100$, and the latent dimensions as $k_1=k_2 = 2$. Furthermore, we set $\beta_1 = \beta_2 = \beta_4 = \beta_5 = 1$ and $\gamma_1 = \gamma_2 = \gamma_4 = \gamma_5 = 1$. For half of the foreground cells, we sampled $t_{j1} \sim \text{Gamma}(1, 1)$ and $t_{j2} \sim \text{Gamma}(1, .01)$. For the other half, we sampled $t_{j1} \sim \text{Gamma}(1, .01)$ and $t_{j2} \sim \text{Gamma}(1, 1)$.

\begin{figure}
\includegraphics[width=1.0\textwidth]{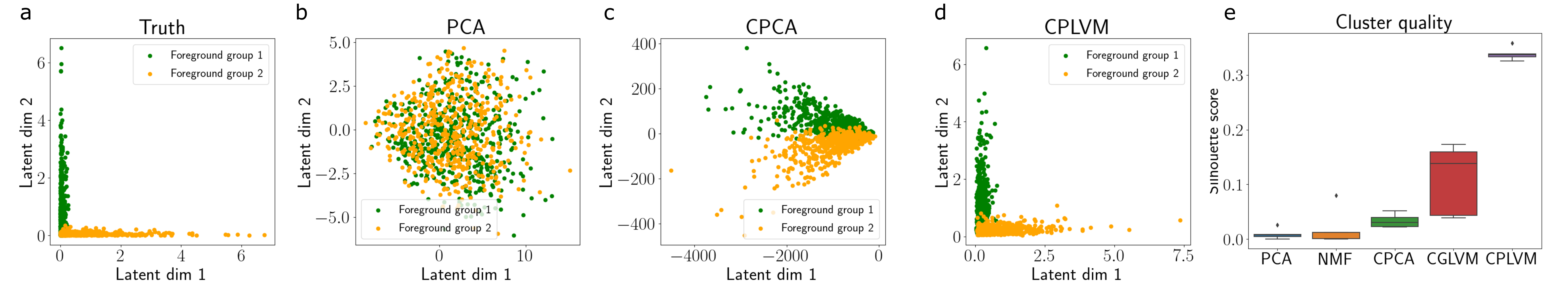}
\caption{\textbf{Cluster identification in simulated data with the CPLVM.} The foreground data was generated from two subgroups of samples. (a) The true underlying foreground-specific latent variables. (b) PCA does not separate the two clusters. (c) CPCA shows an improvement over PCA, but still has overlap in subgroups in the reconstructed data. (d) The CPLVM is able to capture the difference between the subgroups, as well as better preserving nearest-neighbor relationships. (e) Silhouette scores computed on the foreground latent variables for competing methods. }
\label{figure2}
\end{figure}

We fit the CPLVM on this dataset and examine its estimated latent projections of the foreground cells (\autoref{figure2}d). For comparison, we also visualize the latent projections of these cells under PCA and CPCA (\autoref{figure2}b, c). To quantify whether the two foreground subgroups are preserved in the latent space, we compute the silhouette score for the latent variables relative to the true cluster identities. As benchmarks, we also compute the silhouette score for PCA, NMF, CPCA, and CGLVM (\autoref{figure2}e).

We found that the CPLVM's latent space was able to recover the structure of the two subgroups in the foreground (\autoref{figure2}d). In contrast, PCA and CPCA were not able to capture this two-cluster response as clearly (\autoref{figure2}b, c). Moreover, the cluster analysis revealed that the CPLVM was better able to retain the subgroup structure compared to PCA, NMF, CPCA, and CGLVM (\autoref{figure2}e).

\begin{figure}
\includegraphics[width=1.0\textwidth]{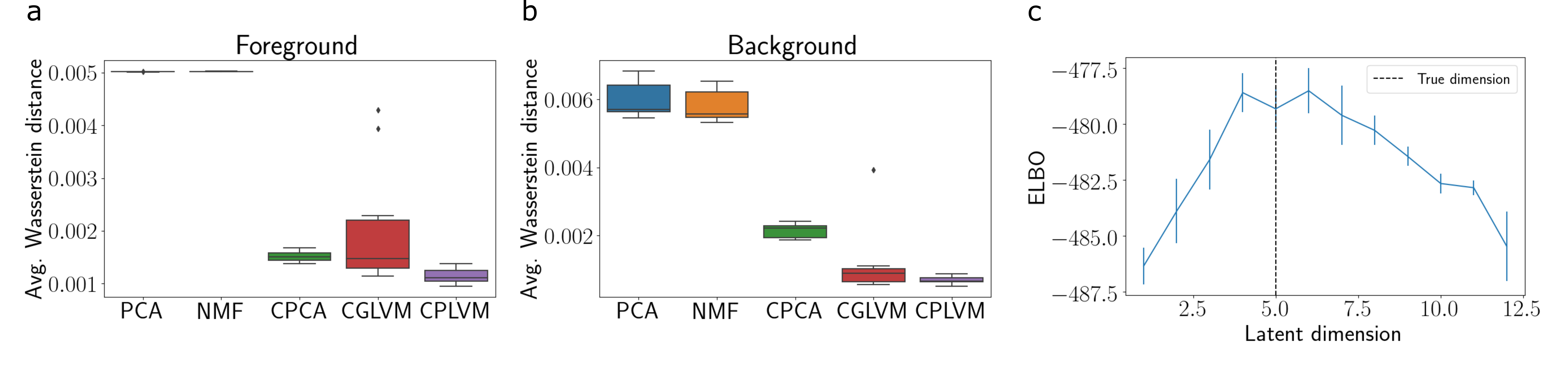}
\caption{\textbf{Simulation experiments with the CPLVM.} We fit our contrastive models to data generated from a small set of shared and foreground-specific latent variables. (a) The average Wasserstein-2 distance between the estimated and true pairwise distances between samples in the foreground for each method. (b) Same as (a), but for background samples. (c) The ELBO for our CPLVM with a range of latent dimensions. The true latent structure of the simulated data is shown by the vertical dotted line. Vertical lines show 95\% confidence intervals.}
\label{figure3}
\end{figure}

To further test the goodness-of-fit of our models, we quantified their ability to recover relationships between samples in their latent spaces. To do this, we generated count data from a small set of latent factors. We then fit four models --- PCA, CPCA, CGLVM, and CPLVM --- and computed the distance between each method's recovered latent variables and the true latent variables. Specifically, we computed the Wasserstein distance between normalized pairwise distance matrices of the simulated and estimated latent variables. We repeated this ten times for each method. We found that the CGLVM and CPLVM outperformed PCA and CPCA in reconstruction error (\autoref{figure3}a, b). The relative performance of the CGLVM and CPLVM was even more noticeable in the background samples, likely due to the CGLVM and CPLVM's explicit models of the background, which PCA and CPCA do not have. In both metrics, the CPLVM outperforms the CGLVM slightly. These results suggest that the CPLVM captures variation in foreground count data relative to background count data, and enables subgroup discovery within the foreground data.


\subsection{Estimating the latent dimension}
Next, we asked whether the CPLVM could estimate the dimension of the generative low-dimensional space. To test this, we generated data from the CPLVM with $k_1 = k_2 = 5$. We then fit a series of CPLVMs, each with a different latent dimension ranging from $k_1 = k_2 = 1$ to $k_1 = k_2 = 15$ with $k_1=k_2$ in all cases. We measured the quality of each model's fit to the data by computing the ELBO for each fit, repeating this procedure ten times for each latent dimension. We found that the ELBO peaked near the true latent dimension $k=5$ (\autoref{figure3}b) and that the ELBO was lower for models with over- or under-constrained latent spaces (corresponding to latent dimensions that were higher or lower than the true dimension, respectively). This result suggests that the CPLVM can recover the true complexity of the variation in the data, and that the ELBO can be used as a reasonable measure of the model's fit to the data. 
% is this true? how does it penalize model complexity? maybe im not thinking of this properly.
% AJ: I was thinking it penalizes complexity in the same way that the evidence does. I.e., since it averages the likelihood over the parameter space, larger models will be penalized. Of course, it's a lower bound here.

\subsection{Hypothesis testing}
Next, we examined whether the CPLVM hypothesis testing framework detects changes in variation between conditions. We consider two types of changes in variation that are found in scientific data: Global shifts in variation across all features, and changes to variation specific to a subset of features~\citep{chandrasekaran2009sparse,leek2008general}.

\subsubsection{Global hypothesis tests}
To evaluate the global hypothesis testing framework, we generated three datasets: one ``alternative'' dataset simulating true global change in variation between conditions and two ``null'' datasets simulating no change between conditions. The alternative dataset --- which we call the \emph{perturbed dataset} --- was drawn from the alternative model defined in \eqref{eq:cplvm1}-\eqref{eq:cplvm3} such that there was substantial change in variation across most genes. The first null dataset --- which we call the \emph{unperturbed null dataset} --- was drawn from the null CPLVM in Equations \eqref{eq:H0_cplvm1}-\eqref{eq:H0_cplvm3} such that there was no change in variation between conditions. The second null dataset --- which we call the \emph{shuffled null dataset} --- was constructed from the samples in the \emph{perturbed dataset} by randomly reassigning cells to the background and foreground conditions.  
These datasets allow us to calibrate the Bayes factors for a truly alternative dataset relative to two truly null datasets. The shuffled dataset is intended to emulate a real-world scenario, in which calibration relative to a true null is not possible.

We computed EBFs for a global hypothesis test for each of the three datasets. For each dataset, we fit the null and alternative models defined in Equations \eqref{eq:H0_cplvm1}-\eqref{eq:H0_cplvm3} and Equations \eqref{eq:cplvm1}-\eqref{eq:cplvm3}, respectively, and computed the EBFs as in Equation \eqref{eq:cplvm_bf}. We repeated this procedure ten times.

\begin{figure}[t!]
    \includegraphics[width=1.0\textwidth]{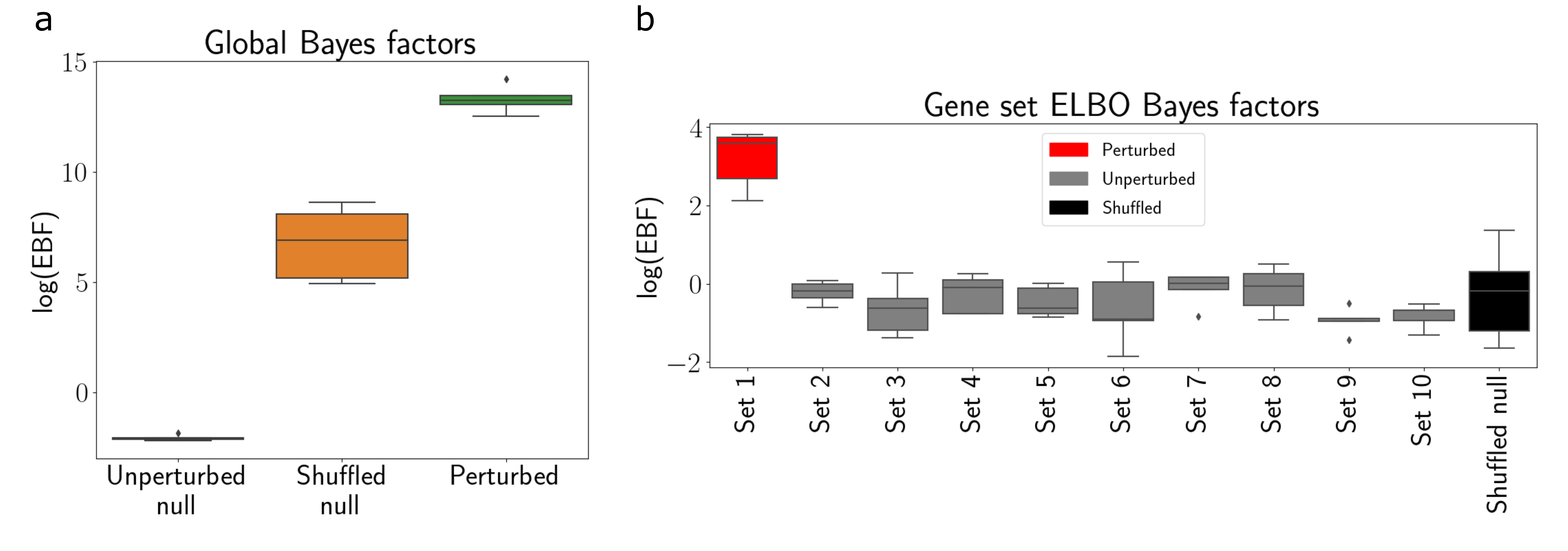}
    \caption{\textbf{Hypothesis testing on simulated data with the CPLVM.} (a) Global hypothesis testing with data generated from a null model (left box), shuffled data approximating truly null data (middle box), and data generated from an alternative model (right box). (b) Gene set hypothesis tests with data in which only variation among genes in gene set one has been altered between conditions (indicated by the red box).}
    \label{figure4}
\end{figure}

We found that the EBFs for the \emph{perturbed dataset} were all substantially above zero. These EBFs were also higher than the EBFs for either of the truly null datasets, indicating a consistently higher lower bound on the model evidence for the alternative model on the \emph{perturbed dataset} (\autoref{figure4}a). The EBFs for the \emph{unperturbed null dataset} were all below zero, implying that the model evidence did not favor the alternative model in this case. The \emph{shuffled null dataset} showed Bayes factors that were between the other two datasets, but distinct from them both. Using the same datasets, we found that the CGLVM was similarly well-calibrated (\autoref{supp_figure1}). This implies that, for global hypothesis testing, the \emph{shuffled null dataset} can be used as the empirical null to calibrate EBFs in practice.

To assess the reliability of the hypothesis testing framework, we quantified how frequently the CPLVM correctly rejected the null hypothesis. To do this, we classified each Bayes factor as ``accept $H_0$'' or ``reject $H_0$'' for a range of thresholds $\tau_1, \dots, \tau_t$, where
\begin{equation}
    \text{Reject $H_0$} \iff \text{ELBO}_{\mathcal{M}_1} - \text{ELBO}_{\mathcal{M}_0} > \tau_i. \label{eq:decisionrule}
\end{equation}
Using this decision rule, we then estimated the true positive rate (TPR) and false positive rate (FPR) for each threshold $\tau_i$:
\begin{align*}
    \text{TPR}_i &= \mathbb{P}(\text{reject $H_0$} | H_1)  \\
    \text{FPR}_i &= \mathbb{P}(\text{reject $H_0$} | H_0).
\end{align*}
To be precise, the \emph{TPR} is the probability of correctly rejecting the null hypothesis (also called the statistical power), and the \emph{FPR} is the probability of incorrectly rejecting the null hypothesis. 

\begin{figure}
    \includegraphics[width=1.0\textwidth]{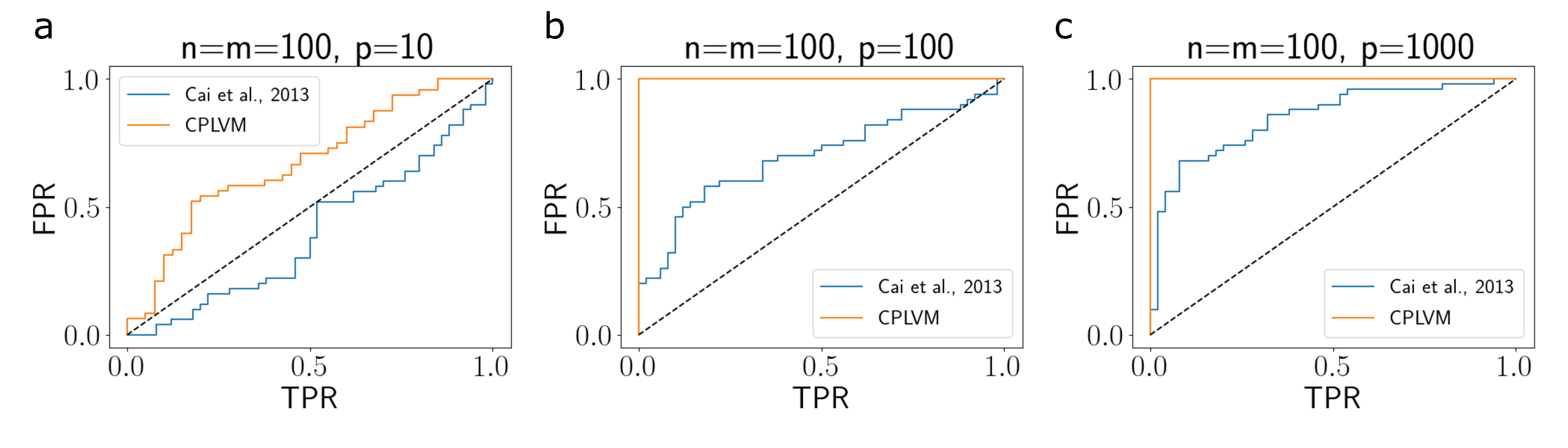}
    \caption{\textbf{Benchmarking the global hypothesis test.} Using simulated data with varying data dimensionalities ($p \in \{10, 100, 1000\}$), we computed ROC curves based on the CPLVM's rejection or acceptance of the null (orange curves). For comparison, we computed the same metrics for Cai's two-sample covariance test that relies on explicitly computing the full sample covariance matrix (blue curves, \citealt{cai2013two}). }
    \label{figure5}
\end{figure}

To quantify how the CPLVM performs under these metrics, we generated data from the CPLVM \eqref{eq:cplvm1}-\eqref{eq:cplvm4}. In particular, we sampled data with three different data dimensions, $p \in \{10, 100, 1000\}$, creating $50$ datasets for each value of $p$. For each setting of $p$, we then computed EBFs for the corresponding datasets. For each dataset, we created a corresponding negative control, or ``null'', dataset by shuffling the foreground or background labels of the samples. Finally, at a range of thresholds $\tau$, we accepted or rejected the null hypothesis for each dataset based on the decision rule in \eqref{eq:decisionrule}. Finally, we computed the TPR and FPR for each value of $\tau$, and we computed the corresponding ROC curves (\autoref{figure5}).

For comparison, we computed the same metrics for a competing two-sample covariance matrix test~\citep{cai2013two}. This approach tests whether the foreground and background covariance matrices are equal,
\begin{equation*}
    H_0: \Sigma_x = \Sigma_y \;\;\;\; \text{versus} \;\;\;\; H_1: \Sigma_x \neq \Sigma_y.
\end{equation*}
This procedure computes a test statistic,
\begin{equation*}
    M_n = \max_{1 \leq i \leq j \leq p} \frac{(\widehat{\sigma}_{kl}^\text{f} - \widehat{\sigma}_{kl}^\text{b})^2}{\widehat{\theta}_{kl}^\text{f} / n + \widehat{\theta}_{kl}^\text{b} / m}
\end{equation*}
where $\widehat{\sigma}_{kl}^\text{f}$ and $\widehat{\sigma}_{kl}^\text{b}$ are the covariance between features $k$ and $l$ in the foreground and background, respectively, and $\widehat{\theta}_{kl}^\text{f}$ and $\widehat{\theta}_{kl}^\text{b}$ are the variance of the covariance elements, 
\begin{align*}
    \widehat{\theta}_{kl}^\text{f} &= \frac{1}{n} \sum\limits_{i=1}^n \left[ (X_{ki} - \Bar{X}_k)(X_{li} - \Bar{X}_l) - \widehat{\sigma}_{kl}^\text{f} \right]^2 \\
    \widehat{\theta}_{kl}^\text{b} &= \frac{1}{m} \sum\limits_{i=1}^n \left[ (Y_{kj} - \Bar{Y}_k)(Y_{lj} - \Bar{Y}_l) - \widehat{\sigma}_{kl}^\text{b} \right]^2.
\end{align*}
Here, $\Bar{X}, \Bar{Y} \in \mathbb{R}^p$ are vectors of sample means. Based on the limiting distribution of $M_n$, the decision rule for this test at level $\alpha$ is
\begin{equation*}
    \text{Reject $H_0$} \iff M_n \geq q_\alpha + 4\log p - \log\log p
\end{equation*}
where $q_\alpha$ is the $1-\alpha$ quantile of the Type I extreme value distribution (Gumbel distribution) with cumulative distribution function $F(x) = \exp\left( -\frac{1}{\sqrt{8 \pi}}  \exp(-x / 2 \right)$.

The ROC curves for these data settings show that the CPLVM test consistently outperforms Cai's two-sample covariance test (\autoref{figure5}). For the CPLVM test, we found that the TPR and FPR remained strong across each data setting, performing perfectly for $p \in \{100, 1000\}$. For the datasets with $p=10$, the CPLVM test did not achieve perfect TPR and FPR, but still performed well above random. In contrast, the two-sample covariance test consistently performed worse than the CPLVM, and indeed performed no better than random guessing with $p=10$ (\autoref{figure5}a). Moreover, the two-sample covariance test had substantially lower TPRs and FPRs than the CPLVM for $p \in \{100, 1000\}$.

These results demonstrate that the CPLVM hypothesis testing framework is able to reliably detect global changes in variation between conditions. Furthermore, the analysis suggests that shuffling cell condition labels is a viable strategy for calibrating the EBFs.

\subsubsection{Gene set hypothesis tests}
To test the gene set hypothesis testing framework, we created ten synthetic genes sets, each made up of $25$ genes. We arbitrarily designated the first gene set --- whose genes are indexed by $\{1, \dots, 25\}$ ---  as the \emph{perturbed gene set}. In other words, this gene set was chosen to show substantial change in variation between the background and foreground conditions. We simulated data for these genes using the CPLVM \eqref{eq:cplvm1}-\eqref{eq:cplvm4}. All other gene sets were designed to not show substantial variation between conditions. We call these truly null gene sets the \emph{unperturbed gene sets}, and we simulated these with the CPLVM corresponding to the null gene set hypothesis (Equations \eqref{eq:cplvm_geneset_H01}-\eqref{eq:cplvm_geneset_H02}). We also included $250$ genes that did not belong to a gene set, which were also simulated from the null gene set model. This led to a total of $p = 500$ genes, half of which belong to gene sets.

To calibrate the gene set EBFs, we estimated an empirical null distribution of EBFs by creating gene sets with randomly assigned genes. In particular, for the $250$ genes belonging to gene sets, we randomly reassigned each of them to synthetic gene sets of size $25$, repeating this $50$ times to create $50$ new synthetic, shuffled gene sets. These gene sets --- which we call \emph{shuffled null gene sets} --- are useful because the true null distribution of EBFs is not available in practice.

For each gene set, we fit the null and alternative models described by \eqref{eq:cplvm_geneset_H01}-\eqref{eq:cplvm_geneset_H02} and \eqref{eq:cplvm1}-\eqref{eq:cplvm3}, respectively, and computed EBFs for each model. We repeated this experiment five times, with each iteration yielding ten EBFs (one for each gene set).

We found that the \emph{perturbed gene set} showed consistently higher EBFs than all other gene sets (\autoref{figure4}b). Furthermore, all EBFs for the \emph{perturbed gene set} were positive, while most other gene sets were consistently negative or near zero. We also found that the EBFs for the \emph{shuffled null} gene sets were also consistently below the \emph{perturbed gene set}, indicating that the EBFs are well-calibrated.

To further test the robustness of the gene set hypothesis tests, we ran the tests in two other simulation settings. First, we tested the robustness of the EBFs to the size of the gene sets. To do this, we generated a similar dataset as before --- containing $500$ genes, $250$ of which belong to gene sets --- but this time we varied the number of genes in each gene set to be in the set $\{1, 5, 10, 15, 20, 25\}$. As expected the EBFs gradually declined when the \emph{perturbed gene set} contained fewer genes (\autoref{supp_figure2}b). However, the test remained robust even for gene sets containing as few as $5$ genes.

Second, we ran the hypothesis test in a setting in which the gene sets were misspecified. In particular, we again constructed gene sets of size $25$, but here only $12$ of the genes in the \emph{perturbed gene set} truly showed a difference between conditions. Even when the gene sets were misspecified as such, we found that the EBFs for the \emph{perturbed gene set} remained substantially above those of the unperturbed gene sets (Supplementary \autoref{supp_figure2}a).

Together, these results imply that the CPLVM gene set hypothesis tests can detect targeted, pathway-specific changes between conditions.

\section{Application to Perturb-seq data}
Next, we applied our models to data from the Perturb-seq platform~\citep{adamson2016multiplexed,dixit2016perturb}.

\subsection{Data}
Perturb-seq is a scRNA-seq platform designed to measure the RNA transcript levels in cells that have been exposed to a set of CRISPR lentivirus guides~\citep{adamson2016multiplexed,dixit2016perturb}. Each guide targets a specific gene, deactivating it by ``cutting'' it out of the genome using a Cas9 nuclease.

In our experiments, for the foreground data, we leveraged Perturb-seq data that contains scRNA-seq measurements on pools of bone marrow-derived dendritic cells (BMDCs), each of which was infected with a unique CRISPR guide~\citep{dixit2016perturb}. Each CRISPR guide in this study was designed to target one of 24 unique transcription factors. For the background data, we use control data from cells that did not receive any treatment. To preprocess the data, for each targeted gene, we pooled data from all CRISPR guides that target that gene. We subsetted each experiment to the 500 most variable genes, according to the Poisson deviance (Supplementary material, \citealt{townes2019feature}).

We fit the CPLVM separately to the datasets from each of the 24 experiments, using the transcript counts from the untreated cells as the background data $\mathbf{Y}$ and the counts from CRISPR-treated cells as the foreground data $\mathbf{X}$.

\subsection{Identifying covariance shifts in Perturb-seq data}
To evaluate the CPLVM's ability to capture shifts in variation in Perturb-seq data, we fit the model for each of the 24 experiments. For comparison, we also fit a Poisson GLM \eqref{eq:poisson_glm1}-\eqref{eq:poisson_glm2} that only identifies changes in the marginal distribution of each gene.

\begin{figure}
\includegraphics[width=1.0\textwidth]{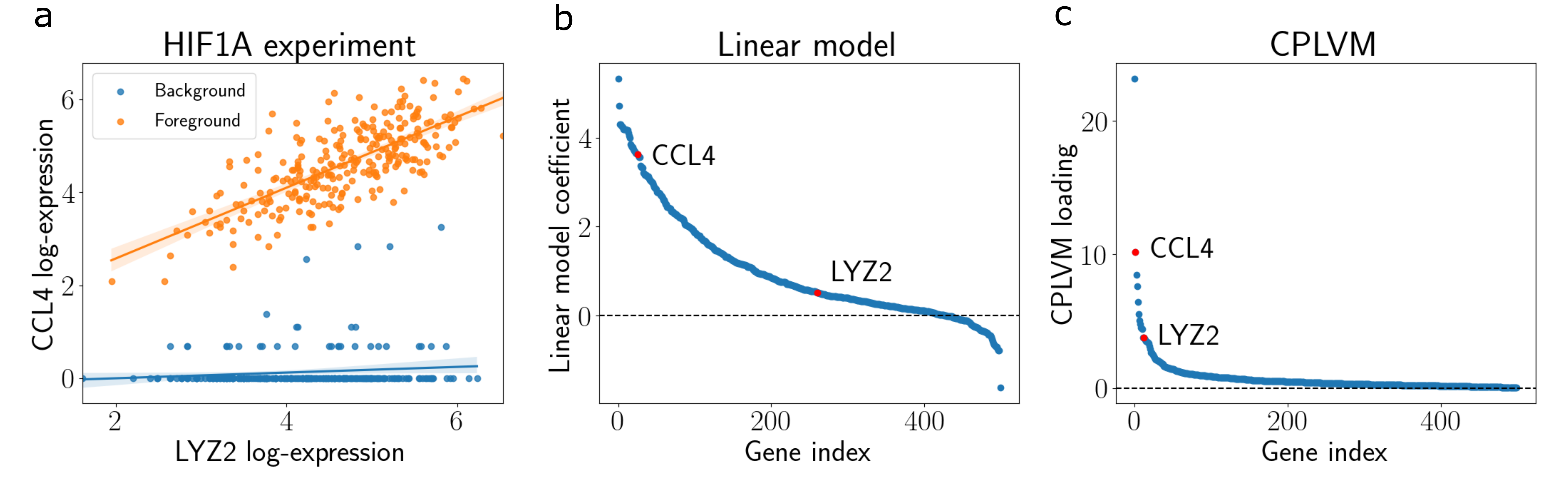}
\caption{\textbf{CPLVM applied to Perturb-seq data.} 
(a) Scatter plot of $\log(x+1)$ expression for \emph{LYZ2} and \emph{CCL4} in the\emph{HIF1A} experiment. \emph{CCL4} shows a positive shift in its marginal expression between conditions, but \emph{LYZ2} does not. However, the correlation between these two genes changes between conditions (Pearson $\rho=0.12$ in the background, and $\rho=0.74$ in the foreground). (b) GLM coefficients for the \emph{HIF1A} experiment. Only \emph{CCL4} is identified as differentially expressed. (c) CPLVM loadings from one CPLVM component for the \emph{HIF1A} experiment. Both \emph{CCL4} and \emph{LYZ2} are identified as having differential variation in this component. }
\label{figure6}
\end{figure}

Examining the CPLVM's latent factors, we found that they identified several shifts in gene-gene covariation that were not picked up by the GLM. One such instance was observed in the \emph{HIF1A}-perturbed experiment. Here, we found that two genes (\emph{LYZ2} and \emph{CCL4}) showed positive correlation across cells in the foreground data, but no correlation in the background data. The CPLVM captured this gene-gene relationship in one of its components (\autoref{figure6}c), while the GLM failed to detect this relationship. Instead, the GLM identified a shift in the marginal expression of \emph{CCL4} alone.

This result suggests that the CPLVM is useful for identifying shifts in variation that occur across multiple genes, and that univariate linear models are not able to detect this type of change.

\subsection{Perturb-seq global hypothesis tests}
Next, we sought to more broadly explore the main sources of variation in each Perturb-seq experiment, and the extent to which each guide induced a substantial change in expression patterns. To do so, we first evaluated the magnitude of the overall change in variation by running global hypothesis tests for each experiment. We computed global EBFs for each (\autoref{figure7}a). To calibrate each test, we also computed EBFs for a second dataset in which cells were randomly reassigned to the foreground or background condition. This shuffled dataset is intended to remove any biologically meaningful patterns that are specific to the foreground data.

Examining the global EBFs, this analysis revealed that most of the experiments showed substantial change in gene expression variation between the untreated and treated conditions. This suggests that most of the guides used in this study had an effect on transcription levels globally across genes, which is expected for these transcription factors. The EBFs for the shuffled datasets were also mostly positive, which was expected from the simulation experiments. However, the EBFs from the shuffled data were consistently lower than their corresponding global EBFs. 

\subsection{Perturb-seq gene set hypothesis tests}
To more narrowly characterize the variation in the Perturb-seq experiments, we performed a series of gene set hypothesis tests. To do this, we leveraged the MSigDB Hallmark gene sets, which categorize genes into a collection of established pathways~\citep{liberzon2015molecular}. For each experiment, we computed the EBF for each Hallmark gene set.

\begin{figure}
\includegraphics[width=1.0\textwidth]{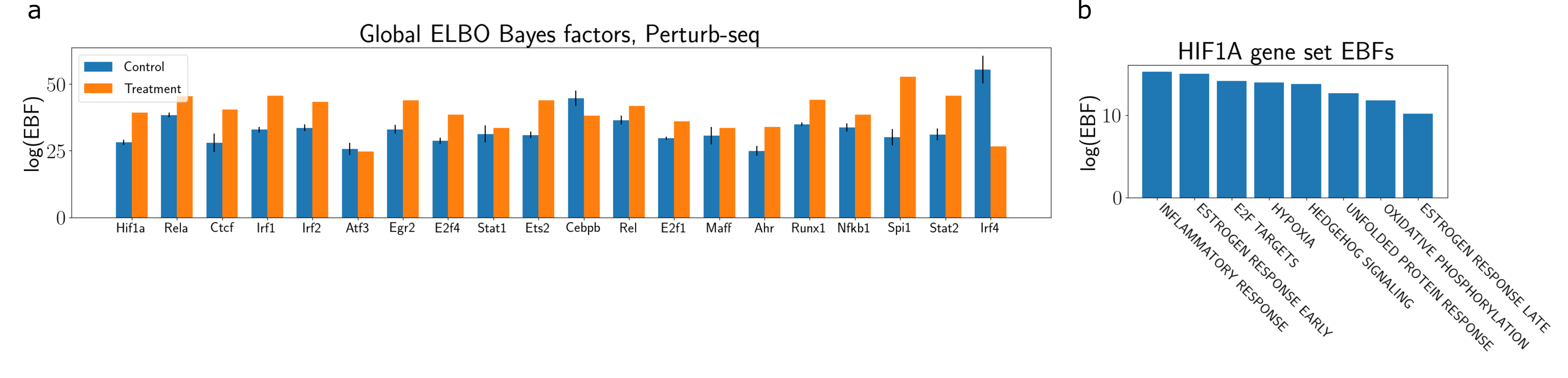}
\caption{ \textbf{Hypothesis testing with Perturb-seq data.} (a) Global hypothesis tests for Perturb-seq experiments. Blue bars represent EBFs for each experiment, and red bars are the EBFs for the shuffled data. Vertical ticks represent 95\% confidence intervals. (b) Gene set EBFs for the \emph{HIF1A} experiment. }
\label{figure7}
\end{figure}

Many gene sets emerged as perturbed from this analysis. For example, in the \emph{HIF1A}-perturbed experiment, a number of coordinated gene sets appeared as top hits, including \emph{TNF-$\alpha$ signaling} and \emph{inflammatory response} (\autoref{figure7}b). Moreover, we found that the magnitude of the gene set EBFs were not correlated with the size of the gene sets (Pearson $\rho=0.03$), suggesting that the tests were not biased by the sizes of the gene sets.
These gene set hypothesis test results suggest that the CPLVM is able to identify coordinated changes in gene expression even among small sets of genes.


\section{Application to small molecule perturbation data}
As further investigation of the CPLVM's behavior on real data, we next applied our model to a scRNA-seq dataset from the MIX-seq platform~\citep{mcfarland2020multiplexed}.

\subsection{Data}
The MIX-seq platform provides scRNA-seq readouts of cancer cell lines' transcriptional responses after being treated with a panel of small molecule therapies~\citep{mcfarland2020multiplexed}. We used a MIX-seq dataset that contains data for 24 cell lines, and we focused on an experiment in which the cells were exposed to idasanutlin, which inhibits the activity of \emph{MDM2}. \emph{MDM2} is known negatively regulate the tumor-suppressor gene \emph{TP53}~\citep{vassilev2004vivo}. Furthermore, idasanutlin has been shown to elicit a selective transcriptional and death response in cells that have wild-type \emph{TP53}, while cells with a mutated copy of this gene do not respond~\citep{mcfarland2020multiplexed}.

\subsection{Application to idasanutlin data}
We fit the CPLVM to the MIX-seq data and analyzed the fitted parameters. We used the transcript counts from idasanutlin-treated cells as the foreground matrix and a the counts from a pool of DMSO-treated cells as the background matrix. In the CPLVM model, we set $k_1=k_2=2$ for visualization.

\begin{figure}
\includegraphics[width=1.0\textwidth]{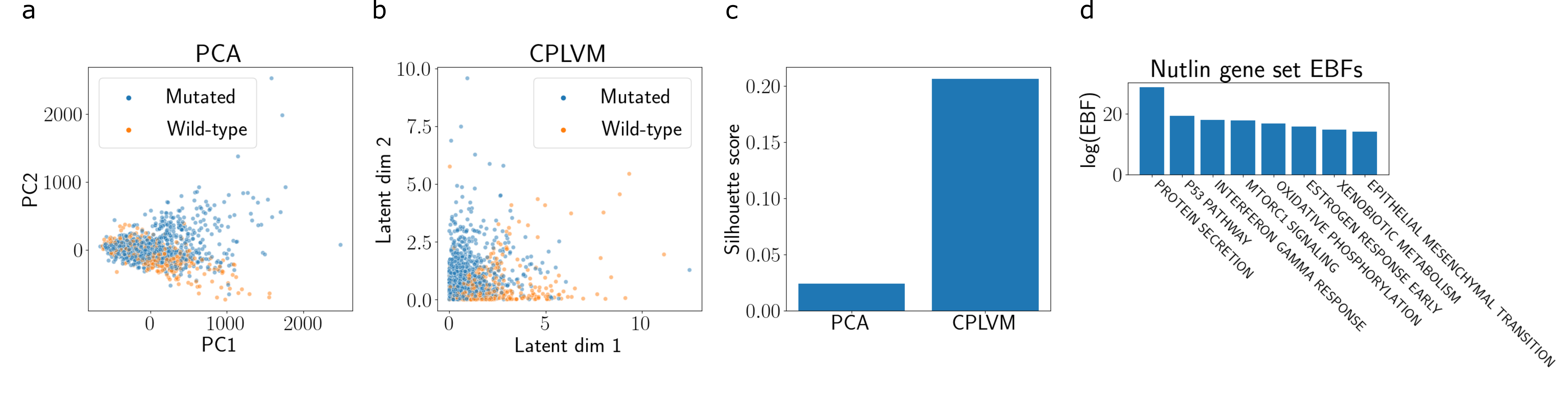}
\caption{\textbf{Contrastive latent variable models applied to chemical perturbation data.} (a) PCA projection of the foreground cells from the idasanutlin experiment. Points (cells) are colored by their \emph{TP53} mutation status. (b) Foreground cells projected into the foreground-specific latent space of the CPLVM. (c) Silhouette score for the clusters of\emph{ }TP53-mutated cells and wild-type cells in the PCA and CPLVM projections. (d) Top gene set EBFs for idasanutlin. The P53 pathway appears as the gene set with the second-highest EBF. }
\label{figure8}
\end{figure}

Visualizing the foreground-specific latent variables for each cell, we found that the CPLVM factors were able to partially separate cells with mutated \emph{TP53} and cells with wild-type \emph{TP53} (\autoref{figure8}b). Meanwhile, a PCA projection of the foreground cells did not clearly identify this subgroup structure (\autoref{figure8}a). A cluster analysis found that the CPLVM latent variables showed tighter clustering of these subgroups compared to PCA (\autoref{figure8}c).

Furthermore, we ran the CPLVM gene set hypothesis tests on the idasanutlin data, again using the MSigDB Hallmark gene sets. This analysis revealed that the \emph{P53 pathway} gene set was among the top enriched pathways (\autoref{figure8}d). This observation coincides with the known mechanism of action of idasanutlin, namely, its direct effect on the \emph{MSM2}/\emph{TP53} pathway~\citep{vassilev2004vivo,mcfarland2020multiplexed}. These results suggests that the CPLVM and corresponding statistical test is able to accurately identify axes of heterogeneity in the response to chemical perturbations, and our model's representation can recover subgroup structure in the foreground data.

\section{Application to GTEx data}
Beyond perturbational data, the CPLVM can be used more generally for count datasets with two conditions. In this section, we demonstrate one such application using bulk RNA-seq data from the Genotype-Tissue Expression (GTEx) Consortium v8 study~\citep{gtex2017genetic,gtex2020gtex}.

The GTEx data contains gene expression measurements in a large number of tissues, collected from thousands of healthy donors. For this experiment, we focused on a subset of the data to answer a specific question: whether there are differences in gene expression variation in coronary artery tissue between donors with and without ischemic heart disease. To do this, we treated gene expression samples from donors with heart disease as the foreground matrix and samples from healthy donors as the background matrix. We subsetted to the 200 most variable genes, and fit the CPLVM on the RNA-seq counts for these samples.

\begin{figure}
\centering
\includegraphics[width=0.8\textwidth]{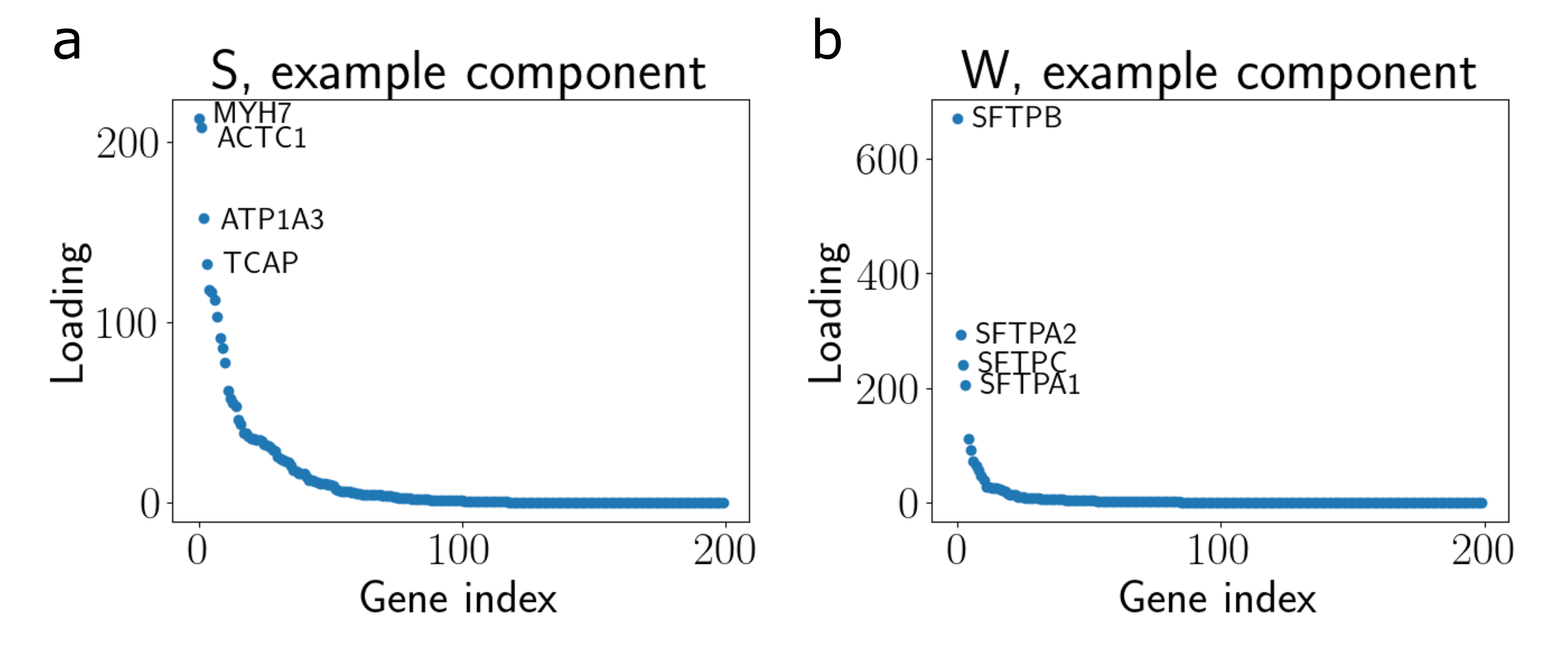}
\caption{\textbf{CPLVM applied to RNA-seq data from coronary artery tissue in patients with and without heart disease.} (a) Sorted loadings values for one component of the shared loadings matrix $\mathbf{S}$. The top genes are related to typical heart function and heart muscle regulation. (b) Sorted loadings values for one component of the foreground-specific loadings matrix $\mathbf{W}$. The top genes are related to oxygen delivery in the heart and lungs, a process that is dysregulated in ischemic artery disease.}
\label{figure9}
\end{figure}

Examining the foreground-specific components of the CPLVM, we found that one of the factors picked up on several genes related to oxygen intake (\autoref{figure9}b). In particular, the genes with the largest coefficients in this factor --- \emph{SFTPB}, \emph{SFTPA2}, \emph{SFTPC}, and \emph{SFTPA1} --- primarily belonged to the pulmonary surfactant protein complex. This complex is known to aid the lung and heart's oxygen-passing abilities.

Furthermore, we examined the parameters of the CPLVM that are shared between the foreground and background samples. We expect these factors to detect variation in expression that exists in both patients with and without heart disease. Indeed, we found that the genes with the highest loading values in one component --- \emph{MYH7}, \emph{DES}, and \emph{MYL2} --- were related to basic heart functioning (\autoref{figure9}a). These results suggest that the CPLVM can be used for settings beyond perturbation experiments, such as for examining structural differences between biological conditions. It also implies that the CPLVM can be used to investigate the shared structure between conditions.

\section{Discussion}
In this study, we presented latent variable models, CPLVM and CGLVM, for case-control sequencing. These models capture the change in gene expression variation that is specific to the case condition, as well as the variation that exists in both the case and control conditions. Our modeling framework provides a set of low-dimensional latent factors that describe this variation. Furthermore, we provide a flexible hypothesis testing framework for characterizing transcriptional structure in case-control experiments. 

Through a series of simulations and experiments with gene expression data, we showed that the CPLVM captures foreground-specific structure and structure that exists in both conditions. In simulations, we showed that that the CPLVM captures the transcriptional variation better than linear models, can identify the proper number of latent dimensions, and enable reliable hypothesis testing of both global and pathway-specific shifts in gene expression. In the context of CRISPR- and drug-treated scRNA-seq data, we showed that CPLVMs can be used to generate biological insights and identify subgroup structure. These insights go beyond traditional differential expression measurements, enabling discovery of differential relationships between genes and cells, such as estimating changes in gene-gene correlations and identifying foreground-specific heterogeneity within a population of cells. 

Several future directions remain to be explored. First, the modeling approach could be extended in several ways. Experiments with more than two conditions could be considered. For example, in gene expression datasets measured across several tissues, it could be useful to model the shared variation among the tissues, as well as the variation that is specific to each tissue~\citep{gtex2020gtex}. Second, more complex inference schemes could be considered. While we used a mean-field variational approximation to the CPLVM, more flexible posterior approximations could be used, such as a variational autoencoder (VAE, \citealt{kingma2014adam,lopez2018deep}). Finally, while our hypothesis testing procedure proved to be well-calibrated, several improvements could be made. Our method uses the ELBO to approximate Bayes factors, but better approximations to the marginal likelihood could be used. Furthermore, our procedure implicitly assigns equal prior weight to both hypotheses, $p(\mathcal{M}_0) = p(\mathcal{M}_1) = 0.5$. This choice proved to be robust in practice, but further investigation into the effect of choosing these priors is warranted.

\section*{Acknowledgements}
The authors thank Danny Simpson and Isabella Grabski for helpful conversations. AJ, FWT, DL, and BEE were supported by a grant from the Helmsley Trust, a grant from the NIH Human Tumor Atlas Research Program, NIH NHLBI R01 HL133218, and NSF CAREER AWD1005627.

\newpage
\bibliographystyle{apalike}
\bibliography{ref.bib}

\begin{thebibliography}{}

\bibitem[Abid et~al., 2017]{abid2017contrastive}
Abid, A., Zhang, M.~J., Bagaria, V.~K., and Zou, J. (2017).
\newblock Contrastive principal component analysis.
\newblock {\em arXiv preprint arXiv:1709.06716}.

\bibitem[Abid et~al., 2018]{abid2018exploring}
Abid, A., Zhang, M.~J., Bagaria, V.~K., and Zou, J. (2018).
\newblock Exploring patterns enriched in a dataset with contrastive principal
  component analysis.
\newblock {\em Nature Communications}, 9(1):1--7.

\bibitem[Adamson et~al., 2016]{adamson2016multiplexed}
Adamson, B., Norman, T.~M., Jost, M., Cho, M.~Y., Nu{\~n}ez, J.~K., Chen, Y.,
  Villalta, J.~E., Gilbert, L.~A., Horlbeck, M.~A., Hein, M.~Y., et~al. (2016).
\newblock A multiplexed single-cell crispr screening platform enables
  systematic dissection of the unfolded protein response.
\newblock {\em Cell}, 167(7):1867--1882.

\bibitem[Anderson, 1958]{lrt}
Anderson, T.~W. (1958).
\newblock {\em An introduction to multivariate statistical analysis}, volume~2.
\newblock Wiley New York.

\bibitem[Aoshima and Yata, 2018]{aoshima2018two}
Aoshima, M. and Yata, K. (2018).
\newblock Two-sample tests for high-dimension, strongly spiked eigenvalue
  models.
\newblock {\em Statistica Sinica}, pages 43--62.

\bibitem[Becht et~al., 2019]{becht2019dimensionality}
Becht, E., McInnes, L., Healy, J., Dutertre, C.-A., Kwok, I.~W., Ng, L.~G.,
  Ginhoux, F., and Newell, E.~W. (2019).
\newblock Dimensionality reduction for visualizing single-cell data using umap.
\newblock {\em Nature Biotechnology}, 37(1):38--44.

\bibitem[Boileau et~al., 2020]{boileau2020exploring}
Boileau, P., Hejazi, N.~S., and Dudoit, S. (2020).
\newblock Exploring high-dimensional biological data with sparse contrastive
  principal component analysis.
\newblock {\em Bioinformatics}, 36(11):3422--3430.

\bibitem[Cai et~al., 2013]{cai2013two}
Cai, T., Liu, W., and Xia, Y. (2013).
\newblock Two-sample covariance matrix testing and support recovery in
  high-dimensional and sparse settings.
\newblock {\em Journal of the American Statistical Association},
  108(501):265--277.

\bibitem[Chandrasekaran et~al., 2009]{chandrasekaran2009sparse}
Chandrasekaran, V., Sanghavi, S., Parrilo, P.~A., and Willsky, A.~S. (2009).
\newblock Sparse and low-rank matrix decompositions.
\newblock {\em IFAC Proceedings Volumes}, 42(10):1493--1498.

\bibitem[Consortium et~al., 2017]{gtex2017genetic}
Consortium, G. et~al. (2017).
\newblock Genetic effects on gene expression across human tissues.
\newblock {\em Nature}, 550(7675):204.

\bibitem[Consortium et~al., 2020]{gtex2020gtex}
Consortium, G. et~al. (2020).
\newblock The gtex consortium atlas of genetic regulatory effects across human
  tissues.
\newblock {\em Science}, 369(6509):1318--1330.

\bibitem[Delmans and Hemberg, 2016]{delmans2016discrete}
Delmans, M. and Hemberg, M. (2016).
\newblock Discrete distributional differential expression (d$^3$e)-a tool for
  gene expression analysis of single-cell rna-seq data.
\newblock {\em BMC Bioinformatics}, 17(1):110.

\bibitem[Dillon et~al., 2017]{dillon2017tensorflow}
Dillon, J.~V., Langmore, I., Tran, D., Brevdo, E., Vasudevan, S., Moore, D.,
  Patton, B., Alemi, A., Hoffman, M., and Saurous, R.~A. (2017).
\newblock Tensorflow distributions.
\newblock {\em arXiv preprint arXiv:1711.10604}.

\bibitem[Ding et~al., 2018]{ding2018interpretable}
Ding, J., Condon, A., and Shah, S.~P. (2018).
\newblock Interpretable dimensionality reduction of single cell transcriptome
  data with deep generative models.
\newblock {\em Nature Communications}, 9(1):1--13.

\bibitem[Dixit et~al., 2016]{dixit2016perturb}
Dixit, A., Parnas, O., Li, B., Chen, J., Fulco, C.~P., Jerby-Arnon, L.,
  Marjanovic, N.~D., Dionne, D., Burks, T., Raychowdhury, R., et~al. (2016).
\newblock Perturb-seq: dissecting molecular circuits with scalable single-cell
  rna profiling of pooled genetic screens.
\newblock {\em Cell}, 167(7):1853--1866.

\bibitem[Finak et~al., 2015]{finak2015mast}
Finak, G., McDavid, A., Yajima, M., Deng, J., Gersuk, V., Shalek, A.~K.,
  Slichter, C.~K., Miller, H.~W., McElrath, M.~J., Prlic, M., et~al. (2015).
\newblock Mast: a flexible statistical framework for assessing transcriptional
  changes and characterizing heterogeneity in single-cell rna sequencing data.
\newblock {\em Genome biology}, 16(1):1--13.

\bibitem[Glass et~al., 2013]{glass2013passing}
Glass, K., Huttenhower, C., Quackenbush, J., and Yuan, G.-C. (2013).
\newblock Passing messages between biological networks to refine predicted
  interactions.
\newblock {\em PloS One}, 8(5):e64832.

\bibitem[Goodman, 1999]{goodman1999toward}
Goodman, S.~N. (1999).
\newblock Toward evidence-based medical statistics. 2: The bayes factor.
\newblock {\em Annals of Internal Medicine}, 130(12):1005--1013.

\bibitem[Hafemeister and Satija, 2019]{hafemeister2019normalization}
Hafemeister, C. and Satija, R. (2019).
\newblock Normalization and variance stabilization of single-cell rna-seq data
  using regularized negative binomial regression.
\newblock {\em Genome Biology}, 20(1):1--15.

\bibitem[Hoffman et~al., 2013]{hoffman2013stochastic}
Hoffman, M.~D., Blei, D.~M., Wang, C., and Paisley, J. (2013).
\newblock Stochastic variational inference.
\newblock {\em The Journal of Machine Learning Research}, 14(1):1303--1347.

\bibitem[Ishii et~al., 2019]{ishii2019equality}
Ishii, A., Yata, K., and Aoshima, M. (2019).
\newblock Equality tests of high-dimensional covariance matrices under the
  strongly spiked eigenvalue model.
\newblock {\em Journal of Statistical Planning and Inference}, 202:99--111.

\bibitem[Kass and Raftery, 1995]{kass1995bayes}
Kass, R.~E. and Raftery, A.~E. (1995).
\newblock Bayes factors.
\newblock {\em Journal of the American Statistical Association},
  90(430):773--795.

\bibitem[Kharchenko et~al., 2014]{kharchenko2014bayesian}
Kharchenko, P.~V., Silberstein, L., and Scadden, D.~T. (2014).
\newblock Bayesian approach to single-cell differential expression analysis.
\newblock {\em Nature Methods}, 11(7):740--742.

\bibitem[Kingma and Ba, 2014]{kingma2014adam}
Kingma, D.~P. and Ba, J. (2014).
\newblock Adam: A method for stochastic optimization.
\newblock {\em arXiv preprint arXiv:1412.6980}.

\bibitem[Kinker et~al., 2020]{kinker2020pan}
Kinker, G.~S., Greenwald, A.~C., Tal, R., Orlova, Z., Cuoco, M.~S., McFarland,
  J.~M., Warren, A., Rodman, C., Roth, J.~A., Bender, S.~A., et~al. (2020).
\newblock Pan-cancer single-cell rna-seq identifies recurring programs of
  cellular heterogeneity.
\newblock {\em Nature Genetics}, 52(11):1208--1218.

\bibitem[Korthauer et~al., 2016]{korthauer2016statistical}
Korthauer, K.~D., Chu, L.-F., Newton, M.~A., Li, Y., Thomson, J., Stewart, R.,
  and Kendziorski, C. (2016).
\newblock A statistical approach for identifying differential distributions in
  single-cell rna-seq experiments.
\newblock {\em Genome Biology}, 17(1):222.

\bibitem[Leek and Storey, 2008]{leek2008general}
Leek, J.~T. and Storey, J.~D. (2008).
\newblock A general framework for multiple testing dependence.
\newblock {\em Proceedings of the National Academy of Sciences},
  105(48):18718--18723.

\bibitem[Li et~al., 2020]{li2020probabilistic}
Li, D., Jones, A., and Engelhardt, B. (2020).
\newblock Probabilistic contrastive principal component analysis.
\newblock {\em arXiv preprint arXiv:2012.07977}.

\bibitem[Li et~al., 2012]{li2012two}
Li, J., Chen, S.~X., et~al. (2012).
\newblock Two sample tests for high-dimensional covariance matrices.
\newblock {\em The Annals of Statistics}, 40(2):908--940.

\bibitem[Liberzon et~al., 2015]{liberzon2015molecular}
Liberzon, A., Birger, C., Thorvaldsd{\'o}ttir, H., Ghandi, M., Mesirov, J.~P.,
  and Tamayo, P. (2015).
\newblock The molecular signatures database hallmark gene set collection.
\newblock {\em Cell Systems}, 1(6):417--425.

\bibitem[Lopez et~al., 2018]{lopez2018deep}
Lopez, R., Regier, J., Cole, M.~B., Jordan, M.~I., and Yosef, N. (2018).
\newblock Deep generative modeling for single-cell transcriptomics.
\newblock {\em Nature Methods}, 15(12):1053--1058.

\bibitem[Love et~al., 2014]{love2014moderated}
Love, M.~I., Huber, W., and Anders, S. (2014).
\newblock Moderated estimation of fold change and dispersion for rna-seq data
  with deseq2.
\newblock {\em Genome Biology}, 15(12):1--21.

\bibitem[McFarland et~al., 2020]{mcfarland2020multiplexed}
McFarland, J.~M., Paolella, B.~R., Warren, A., Geiger-Schuller, K., Shibue, T.,
  Rothberg, M., Kuksenko, O., Colgan, W.~N., Jones, A., Chambers, E., et~al.
  (2020).
\newblock Multiplexed single-cell transcriptional response profiling to define
  cancer vulnerabilities and therapeutic mechanism of action.
\newblock {\em Nature Communications}, 11(1):1--15.

\bibitem[Miao et~al., 2018]{miao2018desingle}
Miao, Z., Deng, K., Wang, X., and Zhang, X. (2018).
\newblock Desingle for detecting three types of differential expression in
  single-cell rna-seq data.
\newblock {\em Bioinformatics}, 34(18):3223--3224.

\bibitem[Nabavi et~al., 2016]{nabavi2016emdomics}
Nabavi, S., Schmolze, D., Maitituoheti, M., Malladi, S., and Beck, A.~H.
  (2016).
\newblock Emdomics: a robust and powerful method for the identification of
  genes differentially expressed between heterogeneous classes.
\newblock {\em Bioinformatics}, 32(4):533--541.

\bibitem[O'Brien, 1992]{o1992robust}
O'Brien, P.~C. (1992).
\newblock Robust procedures for testing equality of covariance matrices.
\newblock {\em Biometrics}, pages 819--827.

\bibitem[Qiu et~al., 2017]{qiu2017single}
Qiu, X., Hill, A., Packer, J., Lin, D., Ma, Y.-A., and Trapnell, C. (2017).
\newblock Single-cell mrna quantification and differential analysis with
  census.
\newblock {\em Nature Methods}, 14(3):309--315.

\bibitem[Robinson et~al., 2010]{robinson2010edger}
Robinson, M.~D., McCarthy, D.~J., and Smyth, G.~K. (2010).
\newblock edger: a bioconductor package for differential expression analysis of
  digital gene expression data.
\newblock {\em Bioinformatics}, 26(1):139--140.

\bibitem[Severson et~al., 2019]{severson2019unsupervised}
Severson, K.~A., Ghosh, S., and Ng, K. (2019).
\newblock Unsupervised learning with contrastive latent variable models.
\newblock In {\em Proceedings of the AAAI Conference on Artificial
  Intelligence}, volume~33, pages 4862--4869.

\bibitem[Srivastava and Yanagihara, 2010]{srivastava2010testing}
Srivastava, M.~S. and Yanagihara, H. (2010).
\newblock Testing the equality of several covariance matrices with fewer
  observations than the dimension.
\newblock {\em Journal of Multivariate Analysis}, 101(6):1319--1329.

\bibitem[Stuart et~al., 2003]{stuart2003gene}
Stuart, J.~M., Segal, E., Koller, D., and Kim, S.~K. (2003).
\newblock A gene-coexpression network for global discovery of conserved genetic
  modules.
\newblock {\em Science}, 302(5643):249--255.

\bibitem[Townes et~al., 2019]{townes2019feature}
Townes, F.~W., Hicks, S.~C., Aryee, M.~J., and Irizarry, R.~A. (2019).
\newblock Feature selection and dimension reduction for single-cell rna-seq
  based on a multinomial model.
\newblock {\em Genome Biology}, 20(1):1--16.

\bibitem[Vassilev et~al., 2004]{vassilev2004vivo}
Vassilev, L.~T., Vu, B.~T., Graves, B., Carvajal, D., Podlaski, F., Filipovic,
  Z., Kong, N., Kammlott, U., Lukacs, C., Klein, C., et~al. (2004).
\newblock In vivo activation of the p53 pathway by small-molecule antagonists
  of mdm2.
\newblock {\em Science}, 303(5659):844--848.

\bibitem[Wang et~al., 2019]{wang2019comparative}
Wang, T., Li, B., Nelson, C.~E., and Nabavi, S. (2019).
\newblock Comparative analysis of differential gene expression analysis tools
  for single-cell rna sequencing data.
\newblock {\em BMC Bioinformatics}, 20(1):40.

\bibitem[Xia et~al., 2015]{xia2015testing}
Xia, Y., Cai, T., and Cai, T.~T. (2015).
\newblock Testing differential networks with applications to the detection of
  gene-gene interactions.
\newblock {\em Biometrika}, 102(2):247--266.

\bibitem[Young et~al., 2018]{young2018single}
Young, M.~D., Mitchell, T.~J., Braga, F. A.~V., Tran, M.~G., Stewart, B.~J.,
  Ferdinand, J.~R., Collord, G., Botting, R.~A., Popescu, D.-M., Loudon, K.~W.,
  et~al. (2018).
\newblock Single-cell transcriptomes from human kidneys reveal the cellular
  identity of renal tumors.
\newblock {\em Science}, 361(6402):594--599.

\bibitem[Zou et~al., 2006]{zou2006sparse}
Zou, H., Hastie, T., and Tibshirani, R. (2006).
\newblock Sparse principal component analysis.
\newblock {\em Journal of Computational and Graphical Statistics},
  15(2):265--286.

\bibitem[Zou et~al., 2013]{zou2013contrastive}
Zou, J.~Y., Hsu, D.~J., Parkes, D.~C., and Adams, R.~P. (2013).
\newblock Contrastive learning using spectral methods.
\newblock {\em Advances in Neural Information Processing Systems},
  26:2238--2246.

\end{thebibliography}

\newpage
\section{Supplementary material}
\subsection{Selecting variable genes}
For all scRNA-seq experiments, we subsetted the data to the most variable 500 genes. We computed the closed-form Poisson deviance for each gene in each dataset using intercept only GLM-PCA as suggested by \cite{townes2019feature}, and took the genes with the highest deviance.

\subsection{Perturb-seq data}
The Perturb-seq data were downloaded from GEO: \url{https://www.ncbi.nlm.nih.gov/geo/query/acc.cgi?acc=GSE90063}.

\subsection{Mix-seq data}
The MIX-seq data were downloaded from Figshare: \url{https://figshare.com/articles/MIX-seq_data/10298696}.

\subsection{Code}
Code for the model and experiments is available at \url{https://github.com/andrewcharlesjones/cplvm}.

\begin{suppfigure}
\centering
    \includegraphics[width=0.5\textwidth]{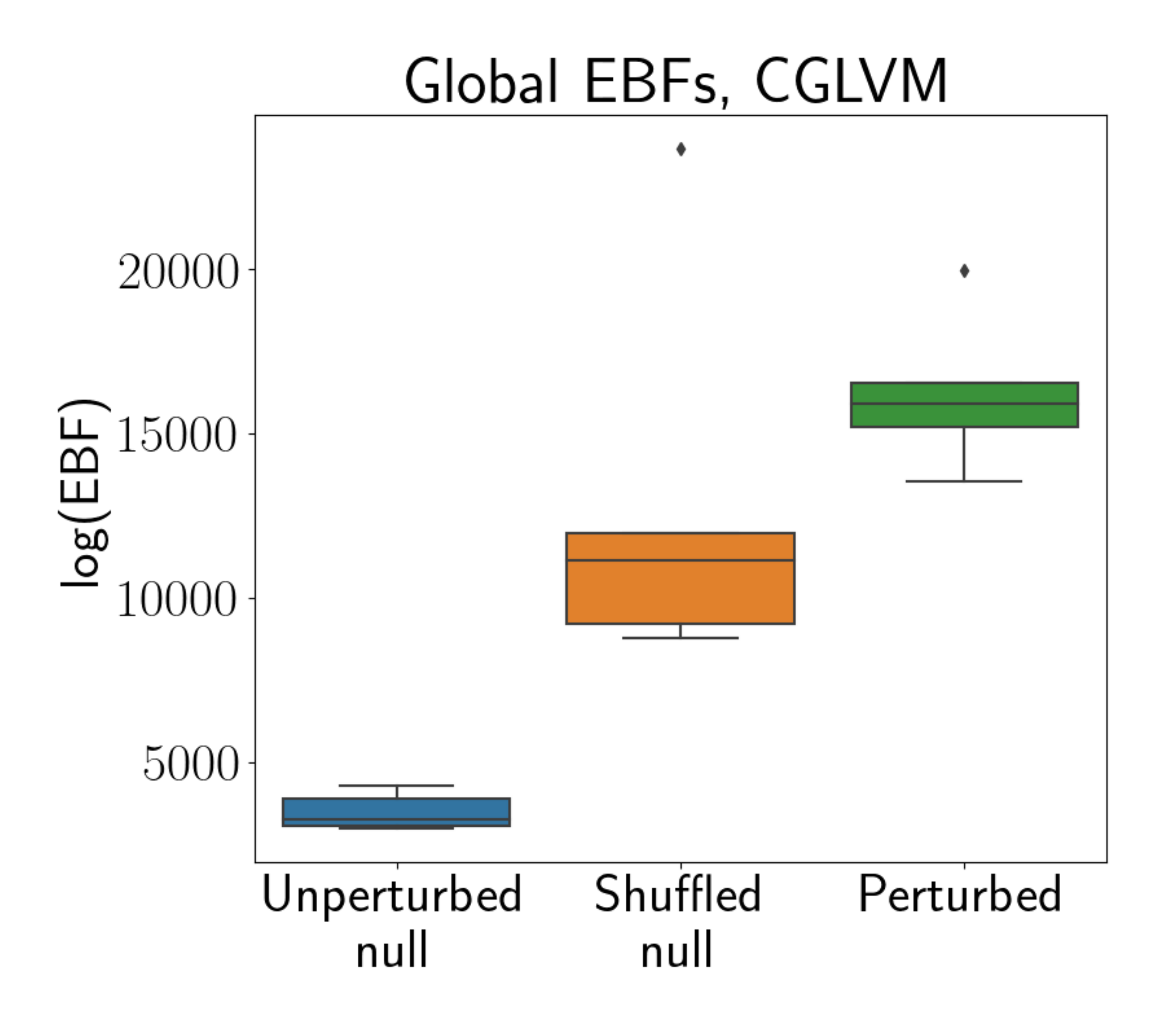}
    \caption{\textbf{Global ELBO Bayes factors for the CGLVM.} Global hypothesis testing using the CGLVM with the same data as used in \autoref{figure4}a. Global tests were run on three datasets: data generated from a null model (left box), shuffled data approximating truly null data (middle box), and data generated from an alternative model (right box). }
    \label{supp_figure1}
\end{suppfigure}

\begin{suppfigure}
    \includegraphics[width=1.0\textwidth]{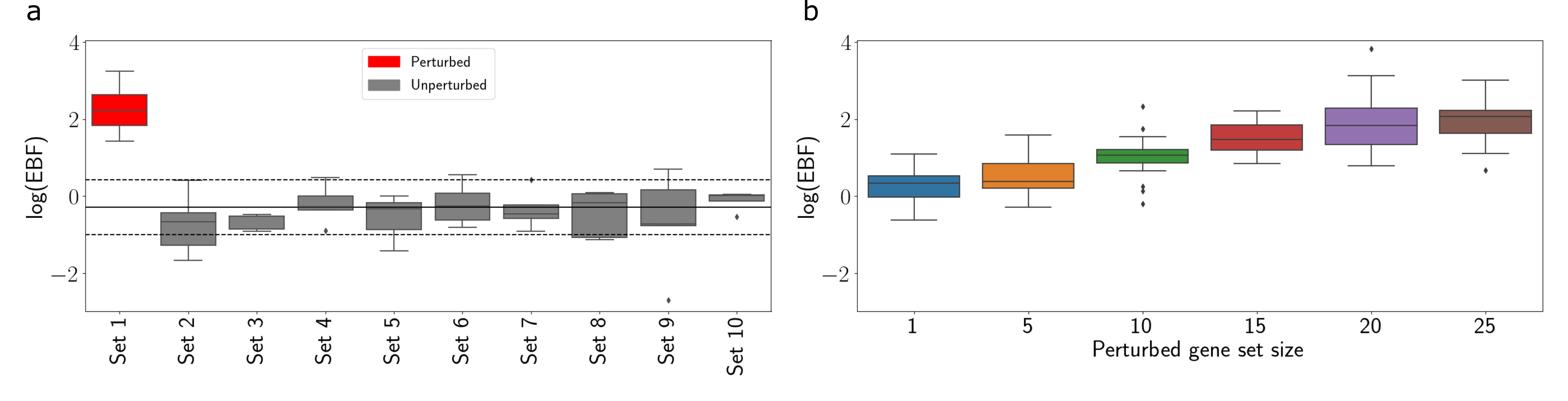}
    \caption{\textbf{Gene set hypothesis tests are robust to gene set misspecification and gene set size.} (a) EBFs for $10$ gene sets, each containing 25 genes. Set 1 was ``perturbed'', but only a fraction (12 of 25) of the genes had substantial differences in variation between conditions. Solid horizontal line shows the mean of the shuffled null, and dotted horizontal lines indicate the 95\% confidence interval for the shuffled null. (b) EBFs for the ``perturbed'' gene set, but at varying gene set sizes. }
    \label{supp_figure2}
\end{suppfigure}

	%%%%%%%%%%%%%%%%%%%%%%%%%%%%%%%%%%%%%%%%%%%%%%%%%%%%%%%%%%%%%%%%%%%%%%
\end{document}